\pdfoutput=1
\documentclass[a4paper,fleqn,usenatbib]{mnras}
\RequirePackage{fixltx2e}

\usepackage{txfonts}
\usepackage{enumerate}

\usepackage{subfigure}

\usepackage[T1]{fontenc}
\usepackage{ae,aecompl}
\usepackage{multicol}

\usepackage{graphicx}	
\usepackage{amssymb}	
\usepackage{wasysym}
\usepackage{color}
\usepackage{xspace}

\newcommand{\text}[1]{\mathrm{#1}}

\newcommand{\Msun}{\ensuremath{\,\mathrm{M}_\odot}\xspace}
\newcommand{\sk}{Sk --$69\,^{\circ}202$}
\title[Binary merger progenitors, SN 1987A]{The quest for blue supergiants: binary merger models for the evolution of the progenitor of SN 1987A }

\author[Menon \& Heger]{
Athira Menon,$^{1}$
Alexander Heger,$^{1,2,3}$
\\
$^{1}$Monash Centre for Astrophysics (MoCA) and School of Physics and Astronomy, Monash University, Clayton, VIC 3800, Australia\\
$^2$School of Physics and Astronomy, University of Minnesota, Minneapolis, MN 55455, U.S.A.\\
$^3$Centre for Nuclear Astrophysics, Shanghai Jiao Tong University, Shanghai, China\\
}

\date{}

\pubyear{2016}

\begin{document}
\label{firstpage}
\pagerange{\pageref{firstpage}--\pageref{lastpage}}
\maketitle

\begin{abstract}

We present the results of a detailed, systematic stellar evolution study of binary mergers for blue supergiant (BSG) progenitors of Type II supernovae. In particular, these are the first evolutionary models that can simultaneously reproduce nearly all observational aspects of the progenitor of SN~1987A, \sk, such as its position in the HR diagram, the enrichment of helium and nitrogen in the triple-ring nebula, and its lifetime before its explosion. The merger model, based on the one proposed by \citet{pods1992a,pods2007}, consists of a main sequence secondary star that dissolves completely in the common envelope of the primary red supergiant at the end of their merger. We empirically explore a large initial parameter space, such as primary masses ($15\,\Msun$, $16\,\Msun$, and $17\,\Msun$), secondary masses ($2\,\Msun$, $3\,\Msun$, ..., $8\,\Msun$) and different depths up to which the secondary penetrates the He core of the primary during the merger. The evolution of the merged star is continued until just before iron-core collapse and the surface properties of the 84 pre-supernova models ($16\,\text{M}_{\odot}-23\,\mathrm{M}_{\odot}$) computed have been made available in this work. Within the parameter space studied, the majority of the pre-supernova models are compact, hot BSGs with effective temperature $>12\,\text{kK}$ and radii of  $30\,\text{R}_{\odot}-70\,\mathrm{R}_{\odot}$ of which six match nearly all the observational properties of Sk --69\,$^{\circ}$202. 

\end{abstract}


\section{Introduction}
\label{intro}
The remarkable supernova SN~1987A that exploded in the Large Magellanic Cloud (LMC) \citep{west1987} is unique in many regards. Although initially classified as a sub-luminous Type II Plateau supernova (Type II-P SN) due to the presence of HI lines in the optical spectrum, its light curve was unusual. The light curve of SN~1987A began to rapidly rebrighten after its initial decline, by a factor of 100 in a few hours as against days for Type II-P supernovae (Type II-P SNe), and at its maximum was only $\sim10\,\%$ as luminous as most Type II-P SNe  \citep{arnett1989,mccray1993}. The dome-shaped light curve \citep{catchpole1988,hamuy1988} indicated that the progenitor was not a typical red supergiant (RSG) ($R\approx500\,\text{R}_{\odot}-1,000\,\text{R}_{\odot}$) as is the case for all observed Type II-P SNe, but a much more compact star. An examination of previous photographic plates of the explosion site confirmed that the progenitor was indeed a compact B3 Ia supergiant named Sk --$69\,^{\circ}202$ \citep{walborn1987,blanco1987}. From its absolute magnitude measurements and by calibrating it against other B3 supergiants in the LMC, the luminosity of Sk --$69\,^{\circ}202$ was deduced to be log\,$(L/\text{L}_{\odot})=5.15-5.45$, with an effective temperature, $T_\text{eff}=15\,\text{kK}-18\,\text{kK}$  \citep{woosley1988a,woosley1988b,arnett1989,walborn1989}. The radius of the star was thus calculated to be $R=\left(3 \pm 1\right) \times 10^{12}$\,cm ($\approx28\,\text{R}_{\odot}-58\,\text{R}_{\odot}$). \citet{barkat1989} deduce a slightly less luminous and possibly cooler progenitor, with log\,$L/\text{L}_{\odot}=4.90 - 5.11$ and $T_\text{eff}=12\,\text{kK}-19\,\text{kK}$.

Another unusual aspect of SN~1987A, is the shape of the circumstellar material nebula ejected by the progenitor before its explosion. It is a complex, triple-ring nebular structure, distributed in an axi-symmetric but extremely non-spherical manner \citep{wampler1990, burrows1995, france2010}. \citet{fransson1989, lundqvist1996} measured $\text{He/H}=0.25 \pm 0.05$ (ratio by number of atoms), in the nebular material but more recent  estimates have lowered this value to $\text{He/H}= 0.17\pm0.06$ \citep{mattila2010} and $\text{He/H}= 0.14\pm0.06$ \citep{france2011}. Nitrogen is also enhanced in the nebula relative to carbon and oxygen; \citet{lundqvist1996} estimated values of $\text{N/C}\sim5\pm2$ and $\text{N/O}\sim1.1\pm0.4$ while \citet{mattila2010} reported N/O$\sim1.5\pm0.7$. Older estimates for these ratios are-- $\text{N/C}\sim8\pm4$ and $\text{N/O}\sim1.6\pm0.8$ \citep{arnett1989}. These enhancements of helium and nitrogen in the nebular material, which was ejected from the envelope of \sk before its explosion, indicate that the star underwent H-burning through the CNO cycle during its evolution \citep{saio1988, fransson1989, sonneborn1997, france2011}.
\citet{panagia1996} found that the outer rings are less enriched in N/C and N/O, by a factor of $\sim3$ than the corresponding values measured in the inner ring, thus concluding that the outer rings may have been ejected 10\,kyr before the inner ring. These results were contested by \citet{crotts2000}, who through a kinematic study, deduced that all three rings were expelled $\sim20$\,kyr before the supernova explosion. \citet{maran2000} further supported this result, through long-slit optical spectroscopic measurements of the CNO abundances of the rings and found no discrepancies between the inner and outer rings, stating that \citet{panagia1996} may not have taken time-dependent line emissions from the outer rings in to account while measuring these abundances.

Possible enhancements in \textit{s}-process elements, Ba and Sr, were also detected in the spectrum during the recombination phase \citep{mazzali1992,mazzali1995}. The dynamical age of the blue supergiant (BSG), measured from the expansion velocity of the inner ring of the nebula,  is estimated to be about 15\,kyr-20\,kyr before explosion \citep{burrows1995, smith1998, crotts2000}. Most of the mass of the nebula resides in the inner ring; the outer rings each weigh $\sim 0.045\,\text{M}_{\odot}$ \citep{lundqvist1996}. The total mass of the nebula is however, uncertain, although estimates range between $0.34\,\text{M}_{\odot}$ \citep{crotts1991} and $1.7\,\text{M}_{\odot}$ \citep{burrows1995,sugerman2005a,sugerman2005b}. It should be borne in mind however, that these estimates are based on an hour glass model while the circumstellar nebula of of SN~1987A is in the form of a triple-ring structure (Phillip Podsiadlowski, private comm.).

Ring-shaped circumstellar nebulae have been found around other BSGs as well, such as MN18 \citep{gvaramadze2015}, SBW1 \citep{smith2013}, HD 168625 \citep{smartt2002}. An object that is considered a more luminous twin of Sk --$69\,^{\circ}202$ of luminosity log\,$(L/\text{L}_{\odot})=5.78-5.90$ \citep{smartt2002,melena2008} is Sheridan 25 (Sher 25) located in the Milky Way. The nebula surrounding this BSG is enhanced in nitrogen \citep{smartt2002,hendry2008} and has a similar hourglass morphology, mass and kinematics to the triple-ring nebula in of SN~1987A \citep{brandner1997a,brandner1997b} suggesting that Sher 25  may also have undergone a similar evolution as Sk --$69\,^{\circ}202$. 

Since the discovery of SN~1987A, eleven more supernovae have been recorded with similar dome-shaped light curves \citep{taddia2013}, suggesting they had BSG progenitors as well. These are collectively classified as Type II-peculiar supernovae (Type II-pec SNe). Light curve models of these SNe indicate that they have compact ($R\leq 70\,\text{R}_{\odot}$) progenitors and are found to be more abundant in low-metallicity environments \citep{taddia2013,pastorello2012}. \citet{pods1992a} theoretically estimated from population synthesis studies with binary merger and accretion models, that $2-8\,\%$ of all massive stars will explode as Type II-pec SNe. According to observations, these are rare events, forming only $1-3\,\%$ of known CCSNe \citep{smartt2009a,kleiser2011,pastorello2012}, while the majority are Type II-P progenitors, i.e., RSGs \citep{smartt2009a}.  This fraction of known Type II-pec SNe could change with more observations from blind and deeper surveys (zPTF, LSST) (Luc Dessart, private comm.).

Presently, the evolution that leads to the formation of a BSG that would explode as a Type II SN is not entirely certain. Single star models include those of low-metallicity \citep{arnett1989}, extreme mass-loss \citep{maeder1987, wood1988}, restricted-convection \citep{woosley1997, langer1991}, helium-enrichment \citep{saio1988} and rapid-rotation \citep{weiss1988,hirschi2004} (see \citep{arnett1989,pods1992b,smartt2009a} for a full review). \citet{barkat1989} performed a paramterised study which showed that the penetration of the convective envelope in the He core (i.e., the H-free core) shrinks the core and dredges up He and N to the surface. Furthermore, the smaller He core relative to the total mass favours a blue solution for the final model.
A similar parameterised study by \citet{petermann2015} for rotating massive stars, demonstrated that models with small He cores could evolve to BSGs. From the luminosity of Sk --$69\,^{\circ}202$ and by fitting light curves from explosions of single star models, the He core mass and envelope mass were estimated to be $M_\mathrm{He\,core}\approx4\,\text{M}_{\odot}-7\,\text{M}_{\odot}$, 
and $M_\text{env}\approx 5\,\text{M}_{\odot}-10\,\text{M}_{\odot}$ respectively \citep{woosley1988b,nomoto1988,woosley1997}. These implied progenitor single stars of main-sequence mass between $14\,\text{M}_{\odot}-20\,\text{M}_{\odot}$ (ignoring mass loss, rotation and overshoot mixing) \citep{woosley1988a,saio1988,arnett1989,smartt2009b}.

The major difficulties in single star models however, are the extreme fine-tuning of parameters required to obtain the transition from red to blue in the HR diagram (HRD) and their inability to reproduce the unusual composition of the circumstellar material. Most importantly, the single star scenario cannot explain the complex geometry of the nebula and how it was ejected about 20\,kyr before explosion. The single star rotating model of \citet{chita2008} does produce this nebular shape, however, the model does not end its life as a BSG.  

An alternative explanation for the BSG progenitor is through interactions in a binary system. Most massive stars are found in binary or multiple systems \citep{popova1982} and of these a substantial fraction (at least $20\,\%-60\,\%$ of stellar systems) are close enough to interact \citep{tutukov1992, eggleton2008, kobulnicky2007, sana2012, sana2013}.
Today, the most widely accepted binary scenario for  Sk --$69\,^{\circ}202$ is that of a merger. Mergers of binary stars are expected to be the end result of $\sim 10\,\%$ of all stars, as indicated by population synthesis studies \citep{pods1992a,pods2006}. The first studies to investigate merger models for \sk are that of \citet{hillebrandt1989,pods1990} and \citet{pods1992a}. The scenario consists of the companion secondary star merges with the primary RSG, via a case B/C mass transfer, and is completely dissolved in the primary's envelope. A BSG model is formed either due to the enrichment of He in the envelope through dredge-up, which lowers the opacity of the surface \citep{hillebrandt1989} or due to the secondary being dumped on the primary, which increases the latter's envelope mass \citep{pods1989,pods1990}.

The merger model of \citet{pods1992a} involves a common envelope phase, wherein the cores of the primary and secondary stars are embedded in the envelope of the primary. Angular momentum transferred from the spiralling-in orbital motion of the cores to the envelope, leading to its spin-up and subsequent mass ejection. This is the first wholesome progenitor model that showed promise in explaining nearly all the observational aspects of SN~1987A \citep{pods1992b}.  The position of the final models in two evolutionary tracks published by these authors, lie where Sk --$69\,^{\circ}202$ was found before it exploded (Fig.~13 in \citealt{pods1992a}). This model was further developed via hydrodynamic simulation studies of the merger \citep{ivanova2002c,ivanovathesis} and the behaviour of the post-merger model \citep{ivanova2002a}, and the formation of the triple-ring nebula from the merger \citep{morris2007,morris2009}. We independently construct our merger model in the present work, guided by the merger scenario of \citet{pods1992a,pods2007} and the results of the studies by \citet{ivanova2002a,ivanova2002b,ivanova2002c,ivanova2003}.

Paraphrasing these works, the evolutionary scenario invoked for Sk --$69\,^{\circ}202$ is as follows.  It begins with a wide binary system of a $15\,\text{M}_{\odot}-20\,\text{M}_{\odot}$ primary and a $1\,\text{M}_{\odot}-5\text{M}_{\odot}$ secondary, with an initial orbital period of greater than 10\,yr.  When the primary evolves to an RSG with a He-depleted core, it transfers mass on a dynamically unstable timescale on the secondary main-sequence star leading to a common envelope (CE) episode, during which the envelope is partially ejected. The secondary star is engulfed by the envelope of the primary and eventually undergoes a merger over  $\sim 100$\,yr \citep{ivanova2002c,ivanova2003}. After thermally adjusting to its structure, the merged star is expected to contract to a rapidly-rotating BSG which sheds additional mass and finally explodes as a Type II-pec SN. The hot and fast wind of the BSG sweeps up the circumstellar material and shapes it to the triple-ring nebular structure we currently observe \citep{chevalier1995,pods2007,morris2007,morris2009}. In  Section~\ref{merger} we provide a more detailed picture of the binary system and its merger.
Aside from other BSGs with ring-shaped circumstellar nebulae, mergers may also be the origin of progenitors of superluminous Type II-P SNe \citep{vanbeveren2013,justham2014} of Type II-bs \citep{folatelli2015}. Mergers may also have played a role in the formation of rapidly rotating B[e] supergiants (such as R4) of super-massive objects such as $\eta$ Carinae \citep{pods2006}.

The evolutionary models presented in this work are the first that can consistently achieve the three confirmed signatures of Sk --$69\,^{\circ}202$: its position as a BSG in the HRD, its lifetime as a BSG before its explosion and its high N/C, N/O and He/H ratios in the surface. The 84 models we have computed in this work are the first pre-supernova (pre-SN) models for H-rich Type II SNe from binary mergers, the majority of which have $T_\mathrm{eff}\geq12\,\mathrm{kK}$ and hence can be used as progenitor models for other Type II-pec SNe. Understanding the evolution of Sk --$69\,^{\circ}202$, can help shed light on the evolution of other hot luminous B-type giants and in a larger scheme, on the late stages of the lives of massive stars.

In order to obtain the best-fit progenitor binary systems for Sk --$69\,^{\circ}202$, we systematically explore a grid of initial binary components and mixing depths during their merger. In the next section, we provide a background of the binary scenario used in this work, based on the works of \citet{pods1992a,pods1992b,pods2006, pods2007}, the 3D simulations of the merger by \citet{ivanova2002c} and their post-merger evolution in 1D as described in \citet{ivanova2002a, ivanova2002b, ivanova2003}.

\subsection{Binary merger scenario}
\label{merger}

The binary system in these works initially consists of a primary star ($M_{1} = 15\,\text{M}_{\odot}-20$\,$\text{M}_{\odot}$) and a secondary ($M_{2} =  1\,\text{M}_{\odot}- 5$\,$\text{M}_{\odot}$) companion star, both on the main sequence, orbiting with an initial period greater than 10\,yr. As the primary approaches core helium depletion, it expands to a red supergiant (RSG)
which consists of a He core (consisting of a CO core + He shell) and convective envelope. It then overflows its Roche lobe and an unstable case C mass transfer ensues from the primary to the secondary, initiating a common envelope (CE) episode that engulfs the secondary. The system now consists of the He core of the primary and the main-sequence secondary inside the convective envelope. Due to viscous drag forces, the secondary spirals in rapidly towards the core and a fraction of the energy released during the orbital decay is transferred to the outer layers of the CE, spinning it up. When the total orbital energy deposited in the envelope becomes comparable to the envelope binding energy,  the envelope expands and ejects some of its mass.
The aspherical outer rings of the nebula may have formed from the mass ejected during this CE phase.

The spiral-in phase ends when the secondary overflows its Roche lobe (at a separation of about $10 \, \mathrm{R}_{\odot}$) and starts a stable mass transfer to the core of the primary, driven by the friction with the envelope in a period of the order of $\sim100$\,yr. H-rich material from the secondary forms a stream during the accretion and penetrates the He core, causing it to shrink in mass. As the secondary mass accretes on the He core, it also gets mixed in the convective envelope of the primary. A fraction of the H-rich secondary mass also penetrates the He core while an equivalent fraction of the He core is dredged up. The region just below the boundary of the He core is hot enough for the CNO cycle to operate and this burns the fresh fuel of hydrogen to helium and nitrogen. The He core mass that is dredged up to the surface is thus enriched in helium and nitrogen. Mass continues to be transferred from the secondary until it finally gets dynamically disrupted and dissolved in the envelope of the primary. 

At the end of the merger, the structure of the star consists of a smaller He core, surrounded by an envelope of homogenous chemical abundances which comprises of the envelope of the RSG primary, mixed with the mass of the secondary star and the material dredged-up from the core. Such a merger, occuring over a timescale of  $\sim100$\,yr, is classified as a `moderate' merger. The remnant will immediately  appear as an RSG out of thermal equilibrium, then contracts continuously towards hotter temperatures and higher luminosities in the HRD. The star thus transitions from the red to the blue part of the HRD and appears as a near-critically rotating BSG, which sheds mass to form the inner ring of the nebula. It is expected that the post-merger star would live as a BSG for about $15\,\text{kyr}-20\,\text{kyr}$ until its explosion \citep{pods2006,heger1998}.

The best-fit hydrodynamic model of \citet{morris2007}, constructed on the basis of the above merger scenario, can successfully reproduce the triple-ring structure of the nebula, with a chosen inner ring mass of $\sim0.4\,\text{M}_{\odot}$ and and an outer ring mass obtained from the merger itself, of $\sim0.02\,\text{M}_{\odot}$ each. The ring mass has not been rigourously constrained and could be higher than current estimates, depending on how much angular momentum is lost from the spiral-in of the secondary (Phillip Podsiadlowski, private comm.).

\subsection{Aims and structure of this work}

We build an `effective-merger' model in 1D using the stellar evolution code {\sc Kepler} and follow the progress of the post-merger star until the onset of core collapse. Our aims are as follows:

\begin{enumerate}[1.]
\item Run simulations over a grid of initial parameter space consisting of primary and secondary masses and the boundary of mixing in the He core during the merger. These are the three major aspects that control the outcome of the merger. 
\item Analyse the distribution of pre-SN models in the HRD and the number ratios of nitrogen to carbon and oxygen (N/C and N/O) and helium to hydrogen (He/H) in the surface; determine how the choice of initial parameters affect the final pre-SN models. 
\item Identify progenitor candidates of SN~1987A that match the observed characteristics of Sk --$69\,^{\circ}202$.
\end{enumerate}

We describe the code employed, the construction of the binary merger scenario and the initial parameters and models in Section~\ref{method}. In Section~\ref{results} we present the nature of the pre-SN models and those that match Sk --$69\,^{\circ}202$. Finally we discuss our results and enlist the conclusions of our study in Section~\ref{conclusions}.

\section{Methodology}
\label{method}

\subsection{The stellar evolution code: \textbf{KEPLER}}

\begin{table*}
\caption{Opacity tables with temperature and density regimes. $\log\,R= \log\,\rho-3\,\left(\log\,T-6\right)$, where $\rho$ and $T$ are in cgs units.}
\label{opacitytable}
\begin{tabular}{lrc}
\hline
Opacity tables & $\log\,\rho$ & $\log\,T$\\
& ($\text{g}\,\text{cm}^{-3}$) & ($\text{K}$)\\
\hline
OPAL 1995 & $-8.0\ldots1.0$ & $3.75\ldots8.70$\\
Conductive opacities & $0.0\ldots7.0$ & $3\ldots9$\\
Low-temperature opacities & $-8.0\ldots1.0$ & $\leq {4}$\\
\hline
\end{tabular}
\end{table*}

Based on the binary merger scenario in Section~\ref{merger}, we use {\sc Kepler}, an implicit one-dimensional hydrodynamics code that can compute stellar evolution models with rotation and nucleosynthesis \citep{heger2000a, heger2005, woosley2002,woosley2007a}. The code uses the Ledoux criterion for convection. Energy generation follows a 19 isotope nuclear reaction network prior to oxygen depletion and a 128 isotope quasi-equilibrium network thereafter. A detailed description of the nuclear reaction rates used for energy generation can be found in \citet{rauscher2002,heger2002}. The physics of rotation in the stellar interior includes angular momentum transport, time-dependent mixing from various rotational instabilities, along with magnetic torques, turbulent viscosities and diffusivities from the dynamo model (please refer to \citealp{heger2000a}, \citealp{heger2000b} and \citealp{heger2005} for more details). Mass loss in the models arise from rotationally modulated winds \citep{heger2000a} and mass-loss prescriptions are as described in \citet{nieuwenhuijzen1990}. The evolution of the model is terminated at the onset of core collapse which is considered to occur when the infall velocity approaches $9\times 10^{7}$\,cm/s.

We recently updated the opacity tables in {\sc Kepler} which oreviously consisted only of Type I OPAL tables (\citealp{iglesias1996} and Achim Weiss, private communication) to include alpha-enhanced Type I OPAL tables, Type II CO-enhanced OPAL tables and conductive opacities (as detailed in \citealp{potekhin2006}).  The opacities (Table~\ref{opacitytable}) and the routines for interpolating them in metallicity, temperature, density, hydrogen mass fractions and enhancements in C, N, O, and Ne were obtained from Boothroyd's homepage (\url{http://www.cita.utoronto.ca/\~boothroy/kappa.html}). Routines to vary opacity from changes in CNO abundances due to nuclear burning were also included. For temperatures lower than $10^{4}$\,K, composition-dependent low-temperature Rosseland mean opacities were computed with {\sc Aesopus} \citep{marigo2009} which includes various sources of atomic, molecular and collision-induced opacities. The routine to interpolate these opacities was provided by Dr. Thomas Constantino \citep{constantino2014}. 

The new opacities are smaller compared to the values obtained from the old tables. The overall effective temperature and luminosity of the pre-SN models increase significantly with these smaller opacities. The role of correct opacities is thus crucial in determining the evolutionary path of the star.

\subsection{Effective-merger model}
\label{model}

\begin{figure*}
\centering\includegraphics[width=\textwidth]{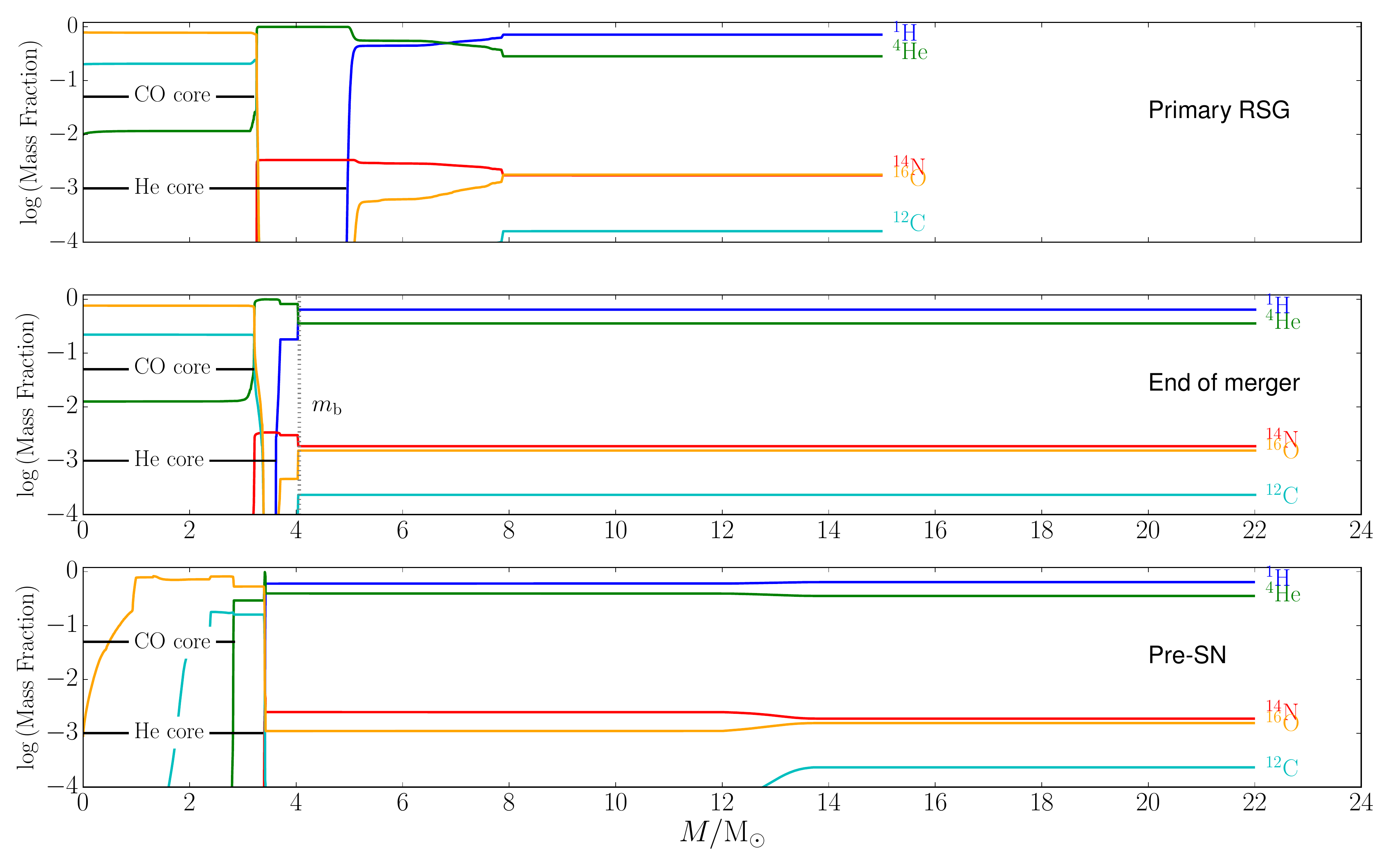}
\caption{Composition of the primary RSG model of $M_{1}=16\,\text{M}_{\odot}$ consisting of a He core of $M_\mathrm{He\,c,1}=4.92\,\text{M}_{\odot}$ just prior to the merger with  $M_{2}=7\,\text{M}_{\odot}$ (top panel, Stage B in Fig.~\ref{hrd1}). The boundary of mixing $m_\text{b}$ (dotted vertical line) is set for $f_\text{c}=16.6\,\%$. In the middle panel, the model plotted is immediately at the end of the merger and has a smaller He core ($3.61\,\text{M}_{\odot}$) (Stage C in Fig.~\ref{hrd1}). The surface composition of the pre-SN model (bottom panel, Stage D in Fig.~\ref{hrd1}) does not change much from the one at the end of the merger. \label{triple_plot}}
~\end{figure*}

\begin{figure*}
\centering\includegraphics[width=0.8\textwidth]{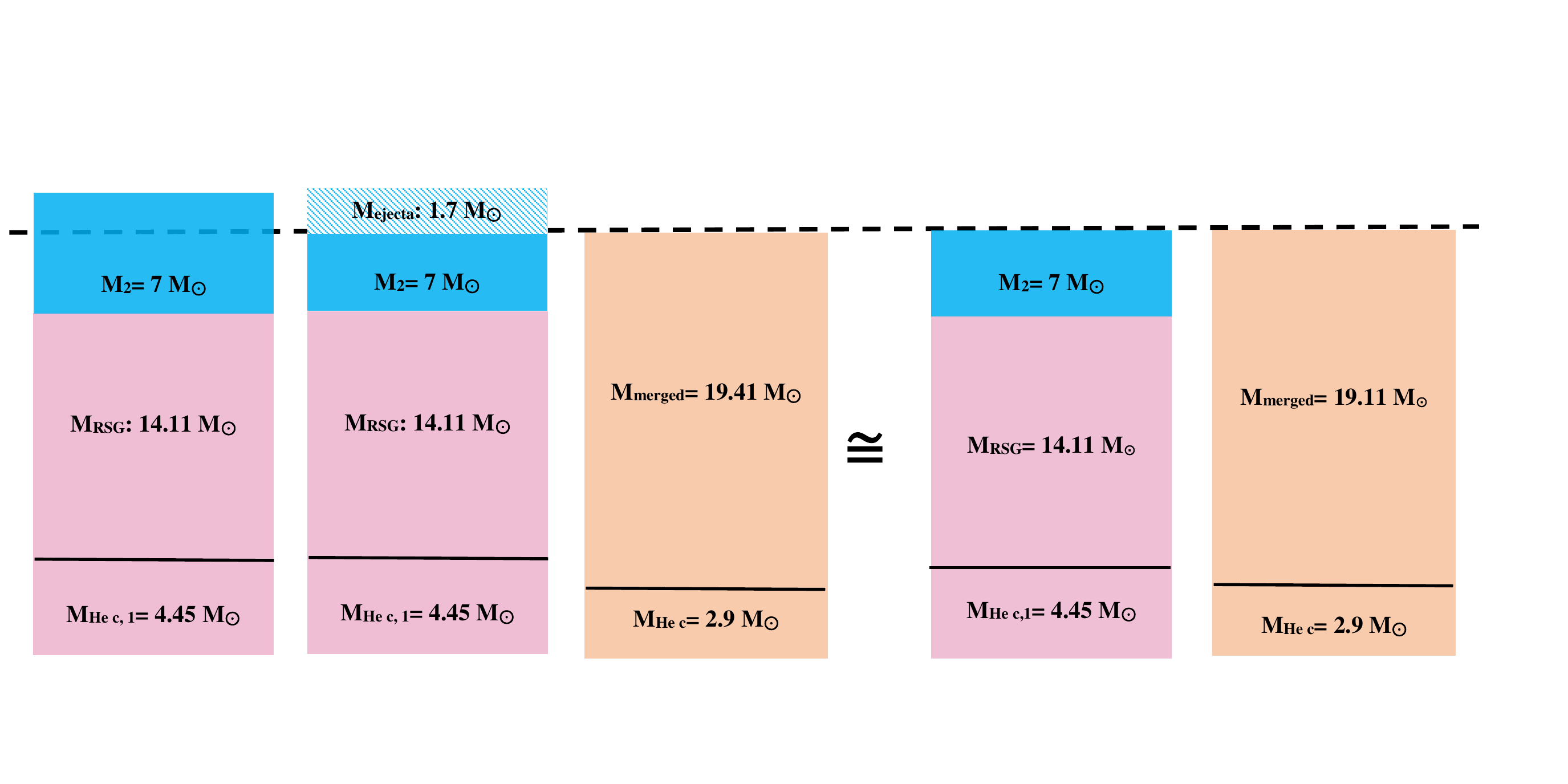}
\caption{Diagram (not according to scale) showing the equivalence of mass ejection from the CE phase to accreting lower secondary masses.}
\label{Mej}
\end{figure*}

\begin{figure*}
\centering\includegraphics[width=0.7\textwidth]{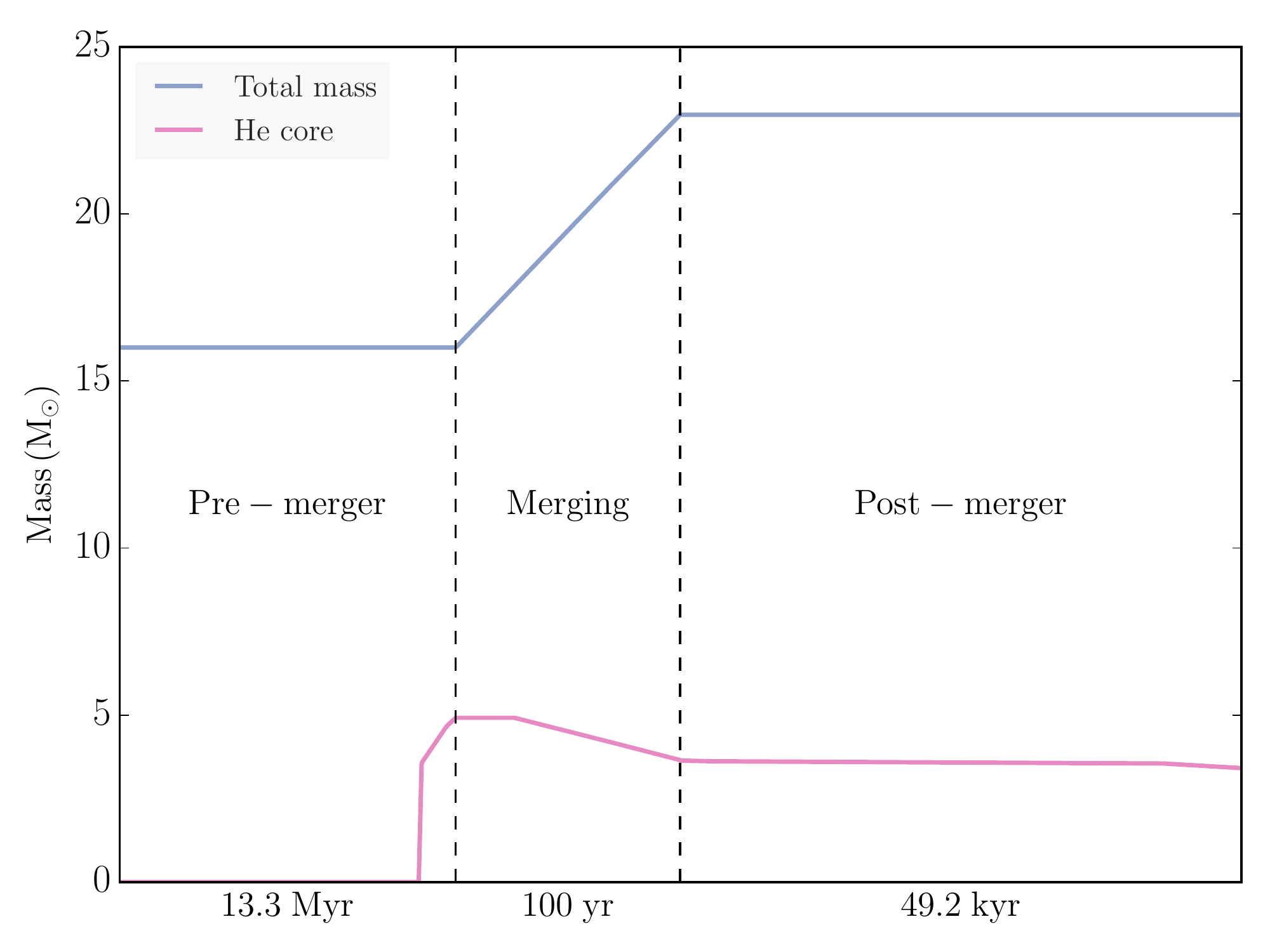}
\caption{A schematic of the merger scenario between $M_{1}=16\,\text{M}_{\odot}$ and $M_{2}=7\,\text{M}_{\odot}$.  $M_{1}$ is evolved until it becomes an RSG, thereafter merging with $M_{2}$. The post-merger remnant remains a BSG for $49.2\,\text{kyr}$ until its explosion. The He core mass is flat until $\sim25$\,yr from the start of the merger before it begins to shrink. This is because it takes $\sim25$\,yr for the boundary of the He core to recede due to dredge up of H-rich material from the envelope.}
\label{schematic}
\end{figure*}

Our 1D effective-merger model is based on the processes outlined in Section~\ref{merger}. The merging phenomenon is characterised by the simultaneous accretion and mixing of the secondary star in the envelope of the primary. 

In this model, we assume merging follows immediately after the primary of main-sequence mass $M_{1}$ evolves to  become the required pre-merger RSG model (as will be described in ~\ref{ini}) whose mass is $M_\text{RSG}$ and consists of a He core of mass $M_\text{He\,c, 1}$. $M_\text{RSG}$ is slightly smaller than $M_{1}$ by $\approx 0.01 \,\text{M}_{\odot}$ due to mass loss through winds. From the hydrodynamic simulations described in Section~\ref{merger}, the merging phase is of the order of 100\,yr and so  in our scenario we choose a fixed merging period of 100\,yr. This merging timescale is much shorter than the thermal time scale of the envelope, some $1000\,$yr, yet the thermal time scale is short compared to the lifetime of the post-merger star before its explosion (of the order of $10^{4}\,$yr-$10^{5}\,$yr), hence varying the merging period within an order of 100\,yr does not affect the post-merger evolution in any significant way. The rate of accretion is $\text{M}_{2}/100\,\text{yr}$, and for the range of ${M}_{2}$ we choose, this leads to accretion rates of $\dot{M}_\text{acc}=0.02\,\text{M}_{\odot}/\text{yr}-0.08\,\text{M}_{\odot}/\text{yr}$. $M_{2}$ is accreted with the same entropy and angular momentum as that of the surface of the primary. 

The merging in our models is implemented as follows. The hydrodynamic simulations show that the secondary star is completely mixed in the convective envelope of the primary. Equivalently we can accrete a secondary star of uniform chemical composition on the primary RSG. In order to obtain this composition, the secondary main-sequence star is evolved to the same age as that of the RSG and then the total masses of individual isotopes are averaged over $M_{2}$. 

As $M_{2}$ gets accreted on the primary, it is also  simultaneously mixed throughout its envelope. This mixing phenomenon is implemented  through a Lagrangian mixing recipe, wherein each unit of $M_{2}$ accreted per timestep of the code ($\dot{M}_\text{acc}\times \text{dt}$) is mixed down progressively in mass to a boundary we specify, $m_\text{b}$, inside $M_\text{He\,c, 1}$. The He core (or the H-free core) mass is defined as the mass within which the hydrogen mass fraction drops to below $\sim 10^{-2}$. As a fraction of $M_{2}$ penetrates the He core of the primary, it brings down H-rich material and thus shrinks the mass of the He core, while an equivalent fraction ($f_\text{c}$) of $M_\text{He\,c, 1}$ is dredged up and mixed uniformly in the envelope. The boundary of penetration or mixing, $m_\text{b}$, of the secondary is thus set by $f_\text{c}$; $m_\text{b}=M_\text{He\,c, 1}-f_\text{c} \times M_\text{He\,c, 1}$. Effectively, $m_\text{b}$ determines the boundary of the He core of the merged star. Since however, a convection zone forms during the merger at $m_\text{b}$, an additional amount of H-rich material is mixed down from the envelope, resulting in a He core boundary that is smaller by $7\,\%-22\,\%$ of $m_\text{b}$.

In this manner, by the end of the merger we have a star that consists of a He core whose mass is smaller by a fraction $\sim f_\text{c}$, and a massive homogenous envelope mixed with $M_{2}$ and $f_\text{c} \times M_\text{He\,c, 1}$ (Fig.~\ref{triple_plot}), resembling the structure of the merged star in the hydrodynamic simulations. The total mass of the post-merger star is $\approx M_\text{RSG}+M_{2}$. 


In this first study, the effective-merger we implement is a simplified model. We do not include any heating of the mass dumped by accretion nor do we track its angular momentum. We also do not compute the angular momentum loss post the CE phase and consequently there is no additional momentum in the envelope or associated mass loss in our models. Our BSG models do not reach break-up velocities after the RSG contracts, hence no mass is shed from the system (aside from the $<0.1\text{M}_{\odot}$ through RSG winds) and we cannot provide predictions for the BSG wind. These processes require detailed hydrodynamic simulations and have been investigated in other works mentioned in Section~\ref{merger}.

Mass loss is, however, an important effect and it does play a significant role in the evolution of a star. Unfortunately, there does not exist an analytical prescription to calculate the mass ejected after the merger \citep{morris2009,vanbeveren2013}. The mass of the circumstellar nebula, currently estimated as $1.7\,\text{M}_{\odot}$, formed from the mass ejected in two stages - during the CE phase and when the post-merger RSG contracted to the BSG. Thus the resultant mass of the star would be smaller by $1.7\,\text{M}_{\odot}$ than the sum of its components $M_\text{RSG}+M_{2}$.

We account for the mass-ejection phenomenon indirectly-- by accreting different values of $M_{2}$ on a particular RSG model. For example, let us take the merger of a system of $M_{1}=15\,\text{M}_{\odot}$ ($M_\text{RSG}=14.11\,\text{M}_{\odot}$) and $M_{2}=7\,\text{M}_{\odot}$, which will result in a star of $21.11 \,\text{M}_{\odot}$. If $1.7\,\text{M}_{\odot}$ is ejected from the merger, this would reduce the total mass to $\approx 19.41 \,\text{M}_{\odot}$. Equivalently, we can merge a system of $M_{1}=15\,\text{M}_{\odot}$ ($M_\text{RSG}=14.11\,\text{M}_{\odot}$) and $M_{2}=5\,\text{M}_{\odot}$ which would result in a star of $19.11\,\text{M}_{\odot}$, close to the mass obtained from the previous system (Fig.~\ref{Mej}). The two systems will also have the same surface composition in the post-merger star. Thus the post-merger evolutionary tracks obtained from both scenarios, the one with mass ejection and the one with a lower $M_{2}$, will be the same.

Fig.~\ref{schematic} outlines the evolutionary sequence for every system-- we begin with the evolution of the primary star from the main sequence, merge it with a secondary mass, and follow the evolution until the post-merger star attains an iron core just prior to core collapse, i.e., the pre-SN model. In the next section we quantify the initial parameters chosen for our study.

\subsection{Initial parameters}
\label{ini}

\begin{table*}
 \caption{Parameters of the primary RSG models selected for merging.  $M_\text{1}$ is the main-sequence mass of primary, ${\omega}/{\omega_\text{crit}}$ is the initial rotation rate on the main sequence; \textsl{Age} is the age of RSG model; $\tau_\text{cc}$ is the time until core collapse; $M_\text{RSG}$ is the mass of RSG model; $M_\text{CO\,c, 1}$ and $M_\text{He\,c, 1}$ are the mass of the CO core and of the He core;  $M_\text{env,1}$ is the mass of the envelope, i.e., $M_\text{RSG}-M_\text{He\,c, 1}$; N/C, N/O, and He/H are surface abundance ratios.}
 \label{primary}
 \begin{tabular}{lcccccccccccc}
  \hline
   $M_\text{1}$ & ${\omega}/{\omega}_\text{crit}$ & Age & $\tau_\text{cc}$ & $M_\text{RSG}$& $M_\text{CO\,c, 1}$& $M_\text{He\,c, 1}$& $M_\text{env, 1}$ & N/C & N/O & He/H\\
   ($\text{M}_{\odot}$) & & (Myr) & (kyr) &($\text{M}_{\odot}$)& ($\text{M}_{\odot}$) & ($\text{M}_{\odot}$) & ($\text{M}_{\odot}$) & & & \\
  \hline
15 & 0.30 & 14.35 & 20.2 & 14.11 & 2.89 & 4.45 & 9.66  & 10.1 & 1.09 & 0.10  \\
16 & 0.30 & 13.28 & 21.2 & 15.04 & 3.25 & 4.87 & 10.17 & 9.2  & 1.12 & 0.10  \\
17 & 0.30 & 12.33 & 37.8 & 15.87 & 3.62 & 5.26 & 10.61 & 9.5  & 0.75 & 0.10 \\
 \hline
\end{tabular}
\end{table*}

\begin{table*}
\caption{Uniform isotopic composition for the accreted secondary model. Isotope: isotopic species, $X_\text{f}$: mass fraction of accreted mass, $X_\text{f}$/$X_\text{i}$: change with respect to initial mass fraction}
\label{secondary}
\begin{tabular}{lcc}
\hline
Isotope & $X_\text{f}$ & $X_\text{f}$/$X_\text{i}$ \\
\hline
$^{1}\text{H}$  & $7.22\times 10^{-1}$ & 0.98\\
$^{4}\text{He}$ & $2.72\times 10^{-1}$ & 1.08\\
$^{12}\text{C}$ & $4.91\times 10^{-4}$ & 0.50\\
$^{14}\text{N}$ & $1.28\times 10^{-4}$ & 4.48\\
$^{16}\text{O}$ & $1.87\times 10^{-3}$ & 0.80\\
\hline
\end{tabular}
\end{table*}

The primary and secondary stars are evolved from the main-sequence with a solar-scaled composition of the LMC: $X_\text{H}=0.739,\, X_\text{He}=0.255$ and $Z=0.0055$, which is 0.4\,dex of the \citet{asplund2009} solar metallicity, $Z_{\odot}=0.014$. This metallicity is the value used by \citet{brott2011}, measured from observations of young massive stars in the H-II regions of the LMC, although they use initial C, N, O values that are enhanced over solar in their work. As we shall discuss later, the metallicity is not the primary reason for stars becoming blue from mergers.

Our choice of main-sequence masses for the binary components is motivated by the mass predicted by single star models for Sk --$69\,^{\circ}202$ and the merger scenario outlined by \citet{pods2007}, i.e, $M_{1}+M_{2}=18\,\text{M}_{\odot}-22\,\text{M}_{\odot}$. 
The primary RSG model chosen for the merger consists of  a convective envelope and a He core with a central helium mass fraction of $\text{X}_\text{He\,c,1}\sim 10^{-2}$ (Fig.~\ref{triple_plot}). The primary main-sequence star has an initial rotation velocity at the equator of ${\omega}/{\omega}_\text{crit}=0.30$ ($v_\text{eq}\sim216$\,km/s). When it arrives on the RSG, its surface is enriched with the ashes of H-burning through the CNO cycle in the core which are dredged up to the envelope due to rotational mixing. Thus the surface of the primary RSG model has high N/C and N/O ratios. Properties of the primary models studied in this work are listed in Table~\ref{primary}.

The main-sequence masses of the secondary considered in this study are between $M_{2}=2\,\text{M}_{\odot}-8\,\text{M}_{\odot}$. Within the age of the primary RSG models ($12.3\,\text{Myr}-14.3\,\text{Myr}$, Table~\ref{primary}), the average isotopic abundances of the secondary masses vary only by a few percent; $X_\text{H}$ ($X_\text{He}$) decreases (increases) by $8\,\%$ between $M_{2}=5\,\text{M}_{\odot}$ and $10\,\text{M}_{\odot}$. This does not significantly impact the evolution of  the post-merger star or its abundances. Hence we choose a `standard' uniform isotopic composition for the accretion of secondary masses-- that of a  $5\,\text{M}_{\odot}$ main-sequence star evolved until $14.3\,\text{Myr}$ (Table~\ref{secondary}).



The initial parameters that we vary between models are:

\begin{enumerate} [1.]
\item \textbf{Primary star mass ($M_{1}$)}: Models of the primary on the main sequence of mass $M_{1}=15\,\text{M}_{\odot}, 16\,\text{M}_{\odot}, 17\,\text{M}_{\odot}$, and initial rotation velocity of ${\omega}/{\omega}_\text{crit}=0.30$ are evolved to the required RSG model (see text above) for the merger.
\item \textbf{Secondary star mass ($M_{2}$)}: Main-sequence star of a mass between $M_{2}=2\,\text{M}_{\odot}-8\,\text{M}_{\odot}$ is merged with each primary RSG model. The initial mass ratio ($M_{2}/M_{1}$) thus spans a range of $0.12-0.53$.
\item \textbf{Fraction of He core dredged up ($f_\text{c}$)}: For each combination of $M_{1}$ and $M_{2}$, we set the boundary of mixing $m_\text{b}$ for $M_{2}$ by specifiying $f_\text{c}$, leading to $m_\text{b}=M_\text{He\,c, 1}-f_\text{c}\times M_\text{He\,c, 1}$. By increasing $f_\text{c}$,  $m_\text{b}$ becomes smaller, resulting in a smaller He core mass for the merged star and larger amounts of $M_\text{He\,c, 1}$ dredged up to the surface. It becomes instructive to use $f_\text{is}$, the fraction of He-shell mass of the He core dredged up, in place of $f_\text{c}$, as we shall see in Section~\ref{results}.
\end{enumerate}

Thus for every model, we choose a value of $M_{1}$ and $M_{2}$ and then choose a value for $f_\text{c}$, which determines $m_\text{b}$. We find that for conditions favourable to form a progenitor model resembling Sk --$69\,^{\circ}202$, $m_\text{b}$ must be on or above the CO core of the primary, as will be explained in Section~\ref{results}. By varying these three parameters, we establish a grid of 84 initial systems to study.

\section{Results}
\label{results}
\subsection{Progenitor models of SN~1987A}
\label{3.1}

A successful SN~1987A progenitor model is one that satisfies the following criteria:
\begin{enumerate}[1.]
\item \label{first}
Lies in the region of the HRD where Sk --$69\,^{\circ}202$ was before exploding;
log\,($L/\text{L}_{\odot})=5.15-5.45$, $T_\text{eff}=15\,\text{kK}-18\,\text{kK}$ and $R/\text{R}_{\odot}=28-58$ \citep{woosley1988a}. 

\item Surface abundances from the BSG model match those of the nebula;
$\text{N/C}\sim5\pm2$ , $\text{N/O}\sim1.1\pm0.4$ \citep{lundqvist1996} and $\text{He/H}= 0.14\pm0.06$ \citep{france2011}. \label{second} 

\item Lifetime of the post-merger BSG phase before explosion is at least $15\,\text{kyr}$. \label{third} 
\end{enumerate}

We classify our pre-SN models as follows:
\begin{enumerate} [$\bullet$]
\item \textbf{BSG}:  $T_\text{eff} \geq 12\,\text{kK}$ 
\item \textbf{YSG}:  $12\,\text{kK} < T_\text{eff} \leq 4\,\text{kK}$ 
\item \textbf{RSG}:   $T_\text{eff} < 4\,\text{kK}$
\end{enumerate}

\begin{figure*}
\centering\includegraphics[width=0.7\textwidth]{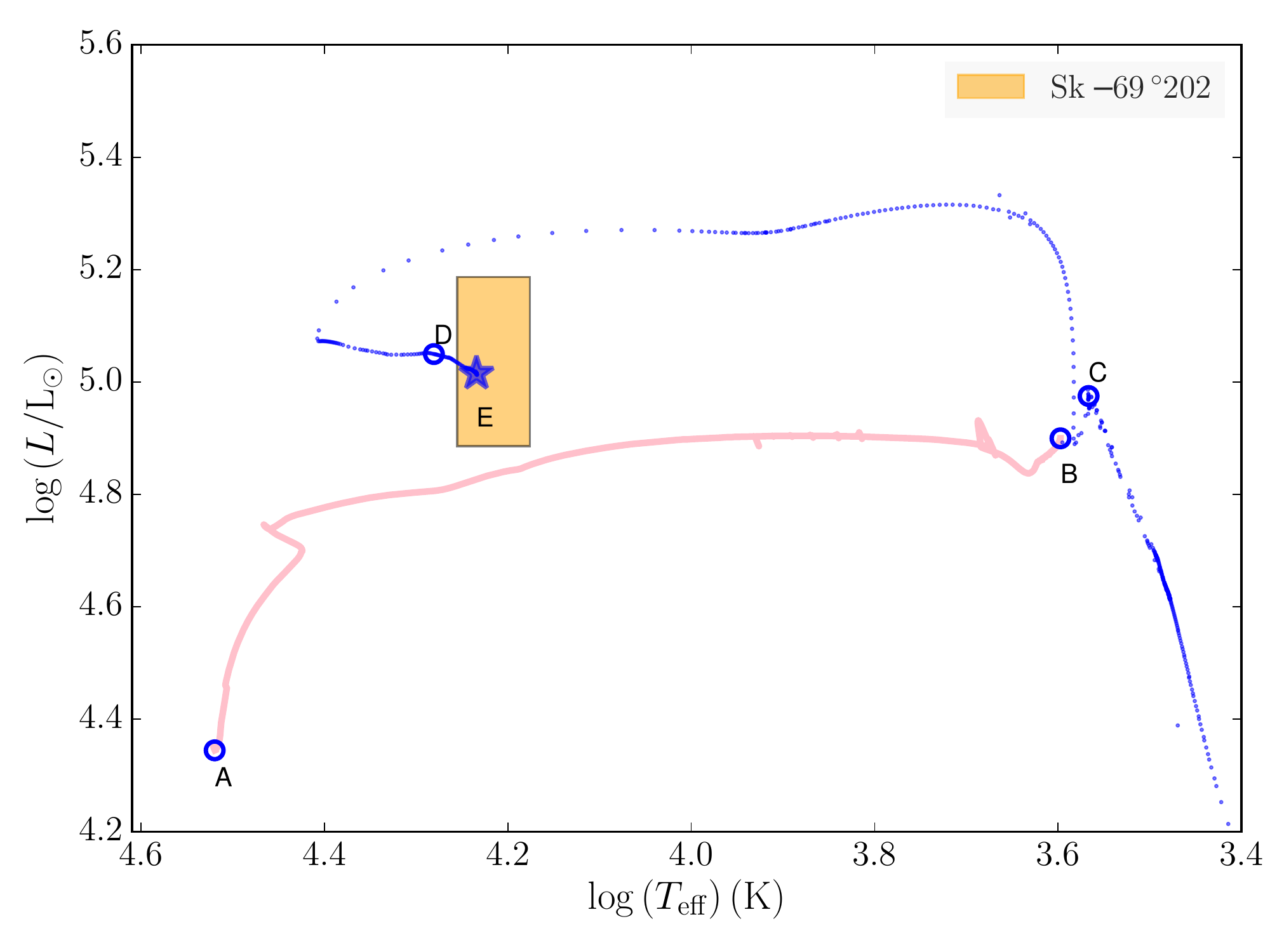}
\caption{Evolutionary track of the merger of $M_{1}=16\,\text{M}_{\odot}$ and $M_{2}=7\,\text{M}_{\odot}$, with $f_\text{c}=16.6\,\%$. Shaded orange region represents observational limits for Sk --$69\,^{\circ}202$ by \citet{woosley1988a}  for $T_\text{eff}$ and log\,($L/\text{L}_{\odot}$). A-B: From zero age main sequence of primary until the required RSG model. B-C: Merger with secondary.  C-D: Evolution of post-merger model until carbon ignition in the core. D-E: Further evolution to final model just before core-collapse. The pre-SN model (E) satisfies the conditions \ref{first}-\ref{third} in Section~\ref{3.1}. 
\label{hrd1}}
\end{figure*}

The evolution of one of the models that resembles the progenitor of SN~1987A ($M_{1}=16\,\text{M}_{\odot}$, $M_{2}=7\,\text{M}_{\odot}$ and $f_\text{c}=16.6\,\%$, see Table~\ref{presn}) is shown in the HRD in Fig.~\ref{hrd1} and in the schematic Fig.~\ref{schematic}. Beginning from the ZAMS model (A), the primary evolves over 13.3\,Myr until helium depletion in the core, during which it inflates to an RSG (B). At this stage, the time until core collapse is 21.1\,kyr. The merger is initiated immediately at point B and the secondary is accreted and mixed with the envelope of the primary until point C over 100\,yr. During the merging process, the star goes out of thermal equilibrium and the code takes small timesteps to evolve it, resulting in a noisy phase on the HRD (the extended dotted blue line in Fig.~\ref{hrd1}). Due to the penetration of H-rich material, the He core mass shrinks, thereby increasing the lifespan of the post-merger star (by nearly 28\,kyr) before it explodes. The H-fuel deposited increases the mass of the H-burning shell and its resulting higher luminosity pushes the convective envelope outward, causing the star to inflate after the merger. When the convective envelope stops expanding and begins to recede, the star contracts and evolves towards the blue part of the HRD. At a certain point in its evolution, the convective envelope stops receding and begins to expand again causing the star to loop back to the red. It then undergoes carbon ignition in the core (D) and subsequent stages of nuclear burning and the evolution is stopped until just before the onset of iron-core collapse (E). The lifespan of this BSG model is 49.2\,kyr before core collapse. 

\begin{table*}
 \caption{Parameters of the pre-SN models for SN~1987A.
$M_\text{1}$ and $M_\text{2}$ are the initial primary and secondary masses of the binary system; $f_\text{sh}$ and $f_\text{c}$ are percent fractions of He-shell mass and helium core mass that were dredged up; $m_\text{b}$ is the boundary of mixing; $M_\text{He\,c}$, $M_\text{Fe\,c}$, $M_\text{env}$, and $M_\text{pre-SN}$ are He core, iron core, envelope masses and mass of the pre-SN model ($M_\text{c}+M_\text{env}$); N/C, N/O, He/H are number ratios of nitrogen to carbon, nitrogen to oxygen, and helium to hydrogen; $T_\text{eff}$, log\,($L$), $R_\text{pre-SN}$ are the effective temperature, luminosity, radius of pre-SN model; $\tau_\text{BSG}$ is the lifetime of the BSG before explosion.}

 \label{presn}
 \begin{tabular}{lrrrrrrrrrrrrrrr}
  \hline
$M_\text{1}$ & $M_\text{2}$ & $f_\text{sh}$ & $f_\text{c}$ & $m_\text{b}$ & $M_\text{He\,c}$ & $M_\text{env}$ & $M_\text{pre-SN}$ & $T_\text{eff}$ & $\log\left(L\right)$  & $R_\text{pre-SN}$ & $M_\text{Fe\,c}$ & N/C & N/O & He/H & $\tau_\text{BSG}$\\

($\text{M}_{\odot}$) & ($\text{M}_{\odot}$) & $\%$ & $\%$ & ($\text{M}_{\odot}$) & ($\text{M}_{\odot}$) & ($\text{M}_{\odot}$) & ($\text{M}_{\odot}$) & (kK) & ($\text{L}_\odot$) & ($\text{R}_{\odot}$) & ($\text{M}_{\odot}$) & & & & (kyr)\\
  \hline
15 & 7 & 50 &17.5 & 3.67 & 2.90 & 18.16 & 21.06 & 16.0 & 4.89 & 2.55 & 1.46 & 6.5 & 1.3 & 0.13 & 82\\
15 & 8 & 50 &17.5 & 3.67 & 2.95 & 19.10 & 22.05 & 17.8 & 4.95 & 2.21 & 1.39 & 5.8 & 1.3 & 0.13 & 83\\
16 & 4 & 10 & 3.30& 4.71 & 4.11 & 14.89 & 19.00 & 16.8 & 4.95 & 2.46 & 1.65 & 6.6 & 1.4 & 0.13 & 41\\
16 & 7 & 50 &16.6 & 4.06 & 3.41 & 18.57 & 21.98 & 17.1 & 5.02 & 2.56 & 0.94 & 6.9 & 1.4 & 0.14 & 49\\
17 & 7 & 10 &15.6 & 4.44 & 3.86 & 18.95 & 22.81 & 16.2 & 5.02 & 2.85 & 1.66 & 7.0 & 1.4 & 0.14 & 41\\
17 & 8 & 10 &15.6 & 4.44 & 3.83 & 19.98 & 23.81 & 17.1 & 5.06 & 2.71 & 1.71 & 6.4 & 1.4 & 0.14 & 41\\

  \hline
 \end{tabular}
\end{table*}

\subsection{What factors affect the formation of BSGs?}

Of the 84 models computed, $59$ are BSGs and $25$ are YSGs. Six of the BSGs qualify as progenitor models of SN~1987A, in accordance with the criteria, \ref{first}-\ref{third} (Fig.~\ref{hrd_rad}, Fig.~\ref{nc_no}) and are summarized in Table.~\ref{presn}. We find that RSG pre-SN models result from mergers only if dredge-up occurs from the envelope, i.e., the He core is not penetrated, as will be  discussed in the following sections. 

We shall now investigate how the criteria in \ref{first}-\ref{third} are affected by the input parameters of our model.

\subsubsection{Surface N/C and N/O ratios}

The envelope of the RSG primary model at the time of merger, is already enhanced in nitrogen at the surface due to rotational mixing, as explained in Section.~\ref{ini} (also see Fig.~\ref{triple_plot}, top panel). Depending on the values of $M_{2}$ and $f_\text{c}$, the N/C and N/O in the envelope will change as explained below.

\begin{figure*}
    \centering
    \begin{subfigure}
        \centering
        \includegraphics[width=0.45\textwidth]{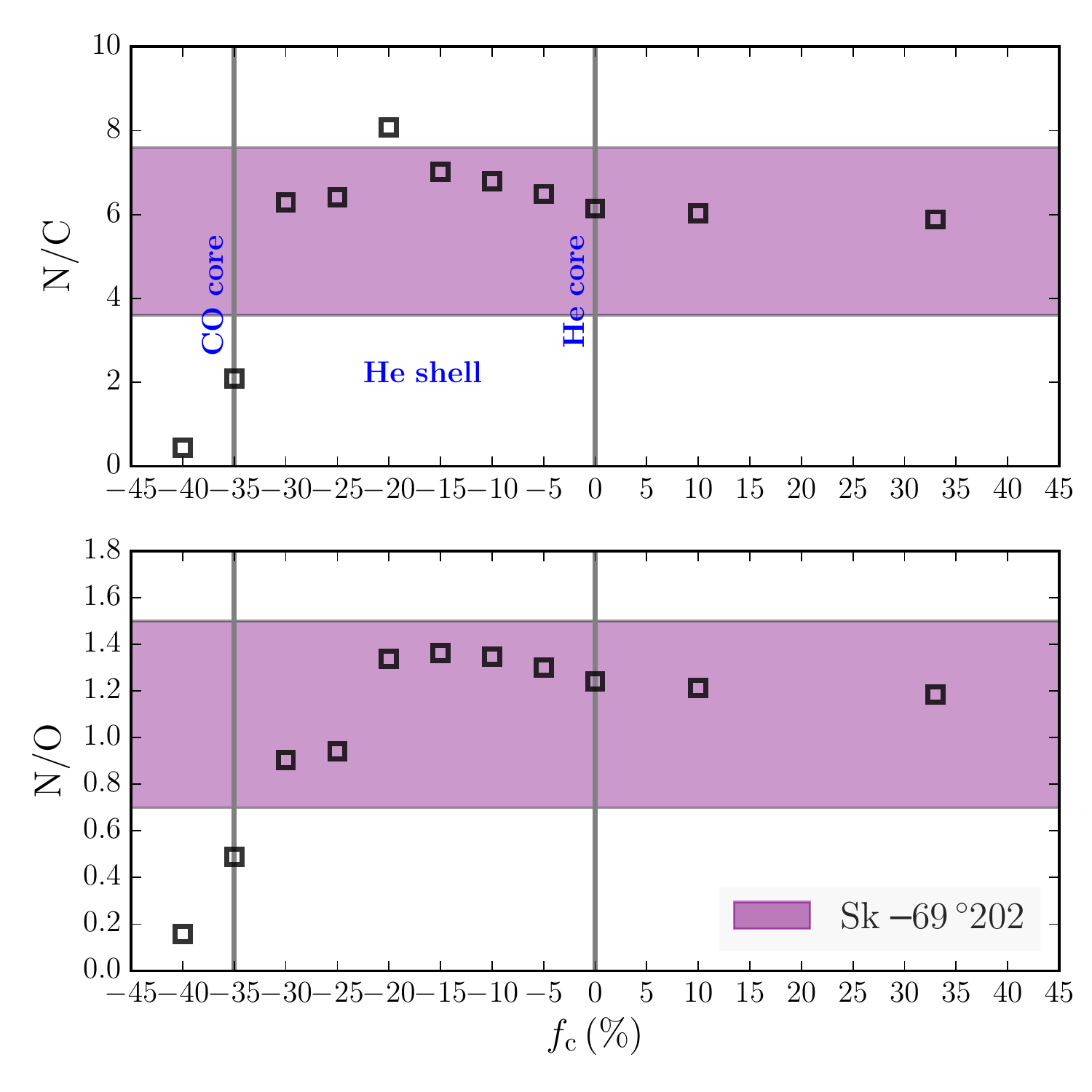}
    \end{subfigure}
    \begin{subfigure}
        \centering
        \includegraphics[width=0.45\textwidth]{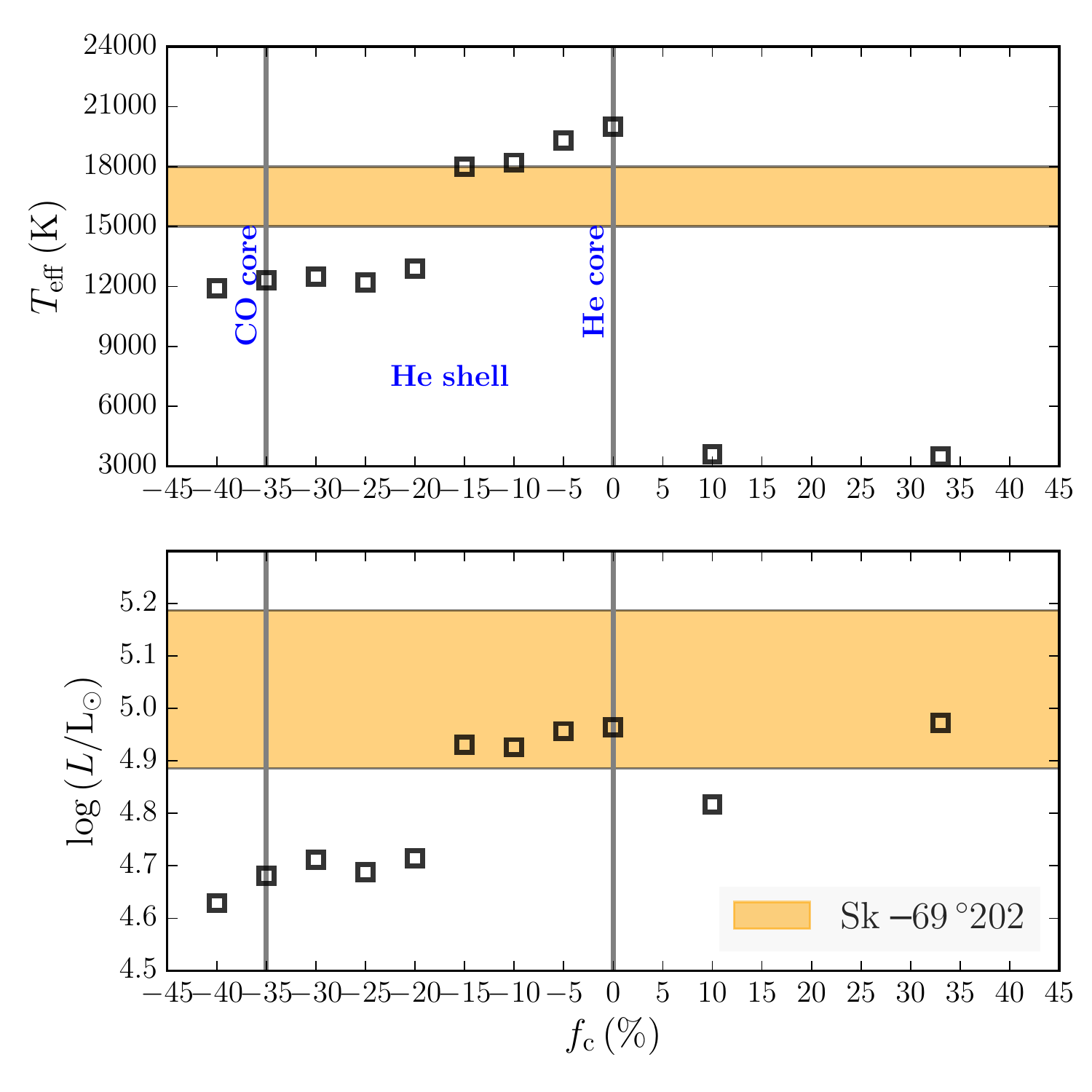}
    \end{subfigure}
    \caption{Surface quantities of pre-SN models obtained from the merger of $M_{1}=15\,\text{M}_{\odot}$ and $M_{2}=5\,\text{M}_{\odot}$ with various dredge up fractions of the He core ($f_\text{c}$ in $\%$).  \textsl{Left:} Number ratios, N/C and N/O.  \textsl{Right:} Effective temperature ($T_\text{eff}$) and luminosity ($L$). 
 Negative values of $f_\text{c}$ represent the case for which the He core of the primary is penetrated and positive values show the case for which the mixing is restricted to above the He core of the primary. Also marked are the boundaries of the CO core, He core and the He shell of the primary. The shaded regions denote observation limits for Sk --$69\,^{\circ}202$; the violet region limits are taken from \citet{lundqvist1996} and the orange region from \citet{woosley1988a}.}
\label{fcore}
\end{figure*}

Our choice for the mixing boundary $m_\text{b}$ being set inside the He core is motivated by two factors -- we know from hydrodynamic simulations that the He core is penetrated by a fraction of the secondary mass during the merger and from our simulations, we find that the surface ratios of N/C and N/O are sensitive to the position of $m_\text{b}$.  In Fig.~\ref{fcore} we demonstrate this for the case of $M_{1}=15\,M_{\odot}$ and $M_{2}=5\,M_{\odot}$ with varying amounts of $f_\text{c}$.  The larger $f_\text{c}$ is, the deeper $m_\text{b}$ is set inside the He core, resulting in larger amounts of $M_\text{He\,c,1}$ being mixed in the envelope during the merger. This is because the He shell, between the boundary of the CO core and He core, is nitrogen rich (Fig.~\ref{triple_plot}). Thus in order to obtain high N/C and N/O ratios at the surface of the merged star, the boundary of  dredge up during the merger must be set within the He-shell region. If this boundary is set within the CO core, the mass dredged up from within the CO core to the surface will be rich in carbon and oxygen, thereby reducing N/C and N/O. Hence it would be more instructive to use $f_\text{sh}$, the fraction of He-shell mass dredged up, to set $m_\text{b}$.

Two models are also computed for the case for which $m_\text{b}$ is set outside the He core of the primary, i.e., there is no He core penetration. For this case, since mass is dredged up from within the homogeneous envelope, the surface values of N/C and N/O are unchanged from their initial amounts.

\begin{figure*}
\centering\includegraphics[width=0.95\textwidth]{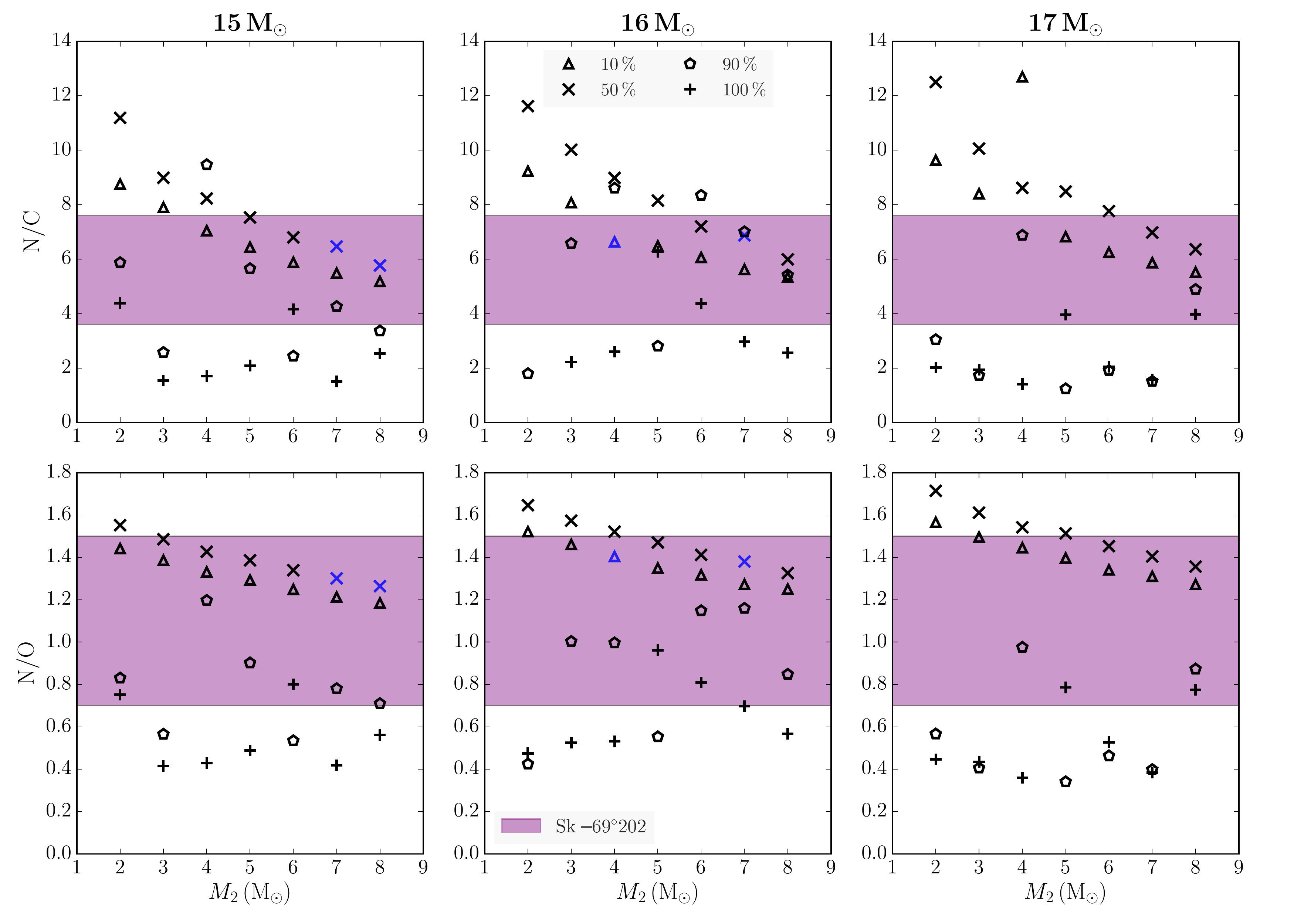}
\caption{Distribution of number ratios, N/C and N/O of all pre-SN models for each of the primary masses, $M_{1}$. These parameters are plotted in a column for each $M_{1}$ against the range of $M_{2}$. The symbols stand for different values of $f_\text{sh}$. The bold blue symbols are progenitor models for SN~1987A that satisfy the criteria \ref{first}-\ref{third} in Section~\ref{3.1}. The shaded violet region denotes the observation limits as explained in Fig.~\ref{fcore}.}
\label{nc-no-m2}
\end{figure*}

We now demonstrate how these quantites vary for all the binary systems studied in this work, spanning the entire initial parameter space of $M_{1}$, $M_{2}$ and $f_\text{sh}=10\,\%$, $50\,\%$, $90\,\%$, and $100\,\%$ (Fig.~\ref{nc-no-m2}). A table containing details of all the pre-SN models computed in this work is provided in Section~\ref{appendix}. N/C and N/O decrease as we dredge up larger fractions of the He-shell material as more carbon and oxygen are dredged up to the surface, as expected from Fig.~\ref{fcore}. As $M_{2}$ increases for a fixed  $f_\text{sh}$, N/C and N/O decrease again. This is because the envelope mass increases as $M_{2}$ increases, causing the amount of nitrogen dredged up to be diluted in the envelope thereby decreasing its mass fraction at the surface. As $M_{1}$ increases, so does the He-core mass (Table~\ref{primary}), thereby increasing the mass dredged up to the surface for a given $f_\text{sh}$. Moreover, the RSG models are increasingly enhanced in N/C and N/O at the surface, hence for a given $M_{2}$ and $f_\text{sh}$, the values of  N/C and N/O at the surface after the merger increase in proportion to $M_{1}$. 

\begin{figure*}
\centering\includegraphics[width=0.7\textwidth]{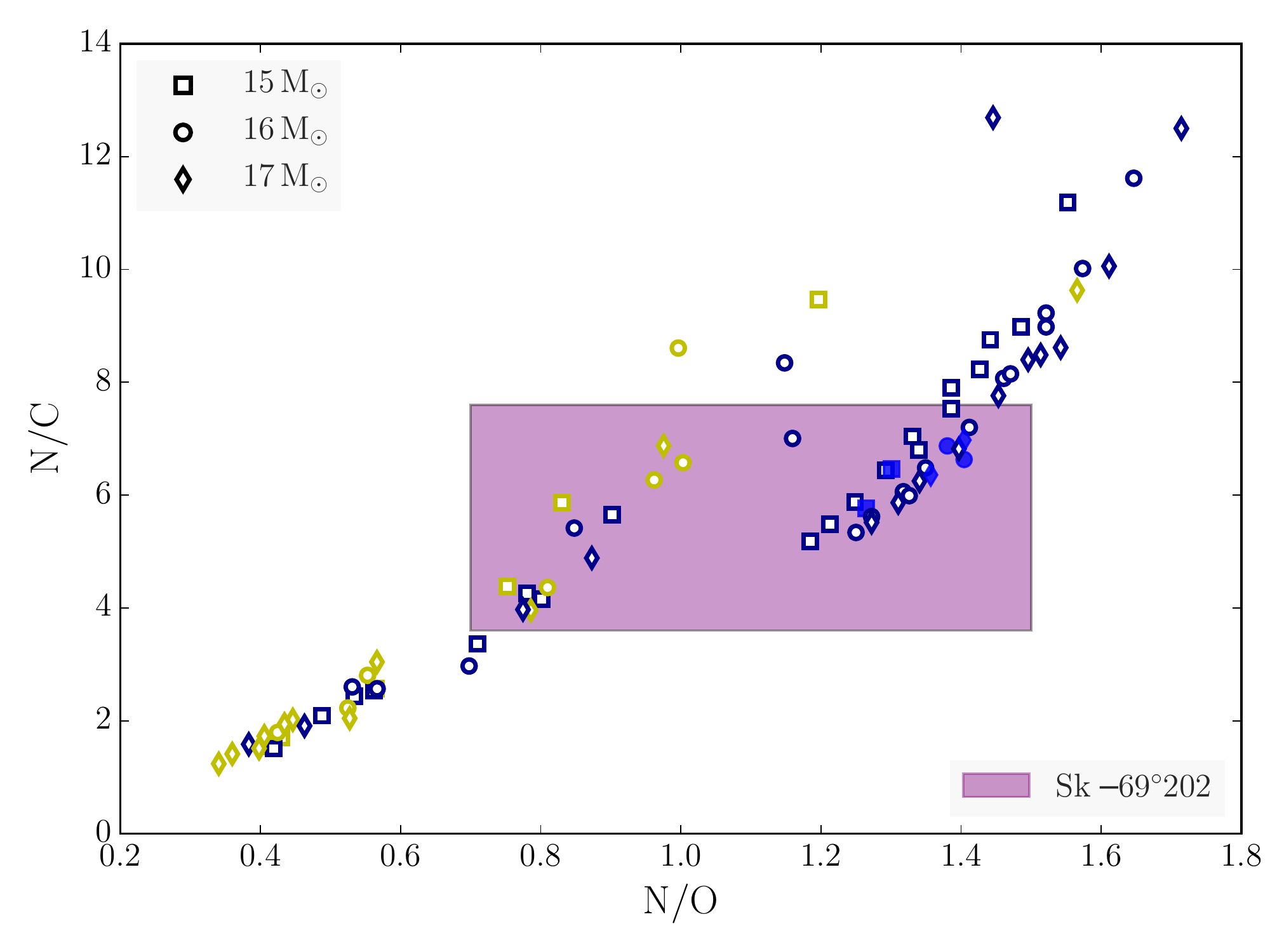}
\caption{Distribution of number ratios N/C vs. N/O at the surface of all 84 pre-SN models computed. The shaded violet region denotes the observation limits for Sk --$69\,^{\circ}202$ as in Fig.~\ref{fcore}. Yellow symbols are YSGs, blue symbols are BSGs and filled blue symbols are progenitor models for SN~1987A, which satisfy the conditions \ref{first}-\ref{third} in Section~\ref{3.1}.}
\label{nc_no}
\end{figure*}

From Fig.~\ref{nc_no} the BSG pre-SN models from our simulations span a large range in surface ratios of N/C and N/O (Fig.~\ref{nc_no}), indicating that there is no correlation between being a BSG and having high values of N/C and N/O at the surface, i.e., these parameters are independent of each other. The YSG models (except for two models) are somewhat more constrained, since they are less enriched in N/C and N/O at the surface than BSGs ($\text{N/O} < 1.0$, $\text{N/C} < 9.7$). The ratio He/H at the surface does not vary significantly for the parameter range we use, and is between 0.13-0.17 for all the pre-SN models. 


\subsubsection{Effective temperature, luminosity, and radius}

\begin{figure*}
\centering\includegraphics[width=0.7\textwidth]{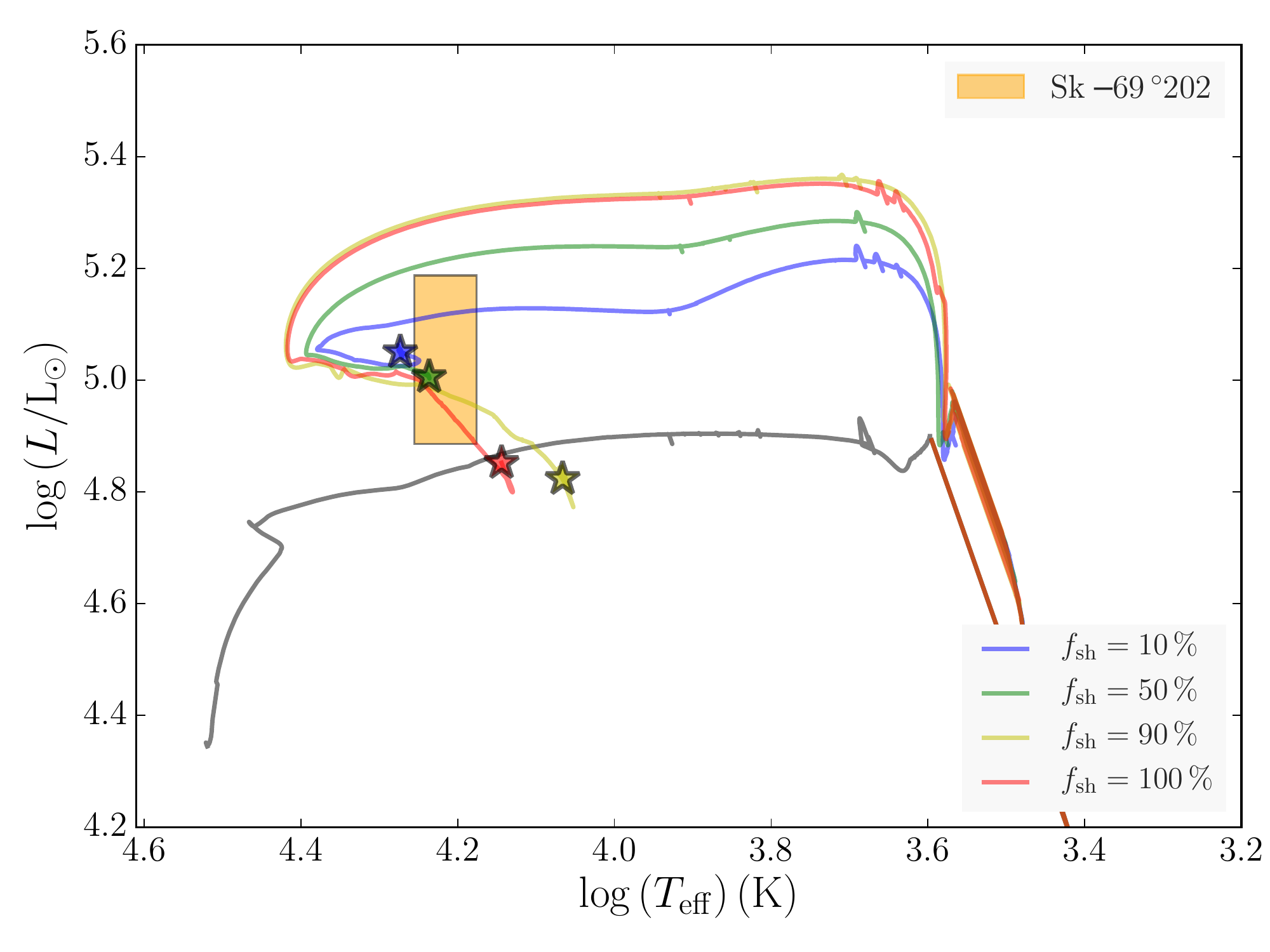}
\caption{HRD of four tracks computed for the merger of $M_{1}=16\,\text{M}_{\odot}$ and $M_{2}=6\,\text{M}_{\odot}$, with $f_\text{sh}=10\,\%$, $50\,\%$, $90\,\%$, $100\,\%$.  Stars denote the pre-SN models of individual evolutionary tracks. The shaded orange region denotes the observation limits as in Fig.~\ref{fcore}. 
\label{hrd_f}}
\end{figure*}

\begin{figure*}
\centering\includegraphics[width=0.7\textwidth]{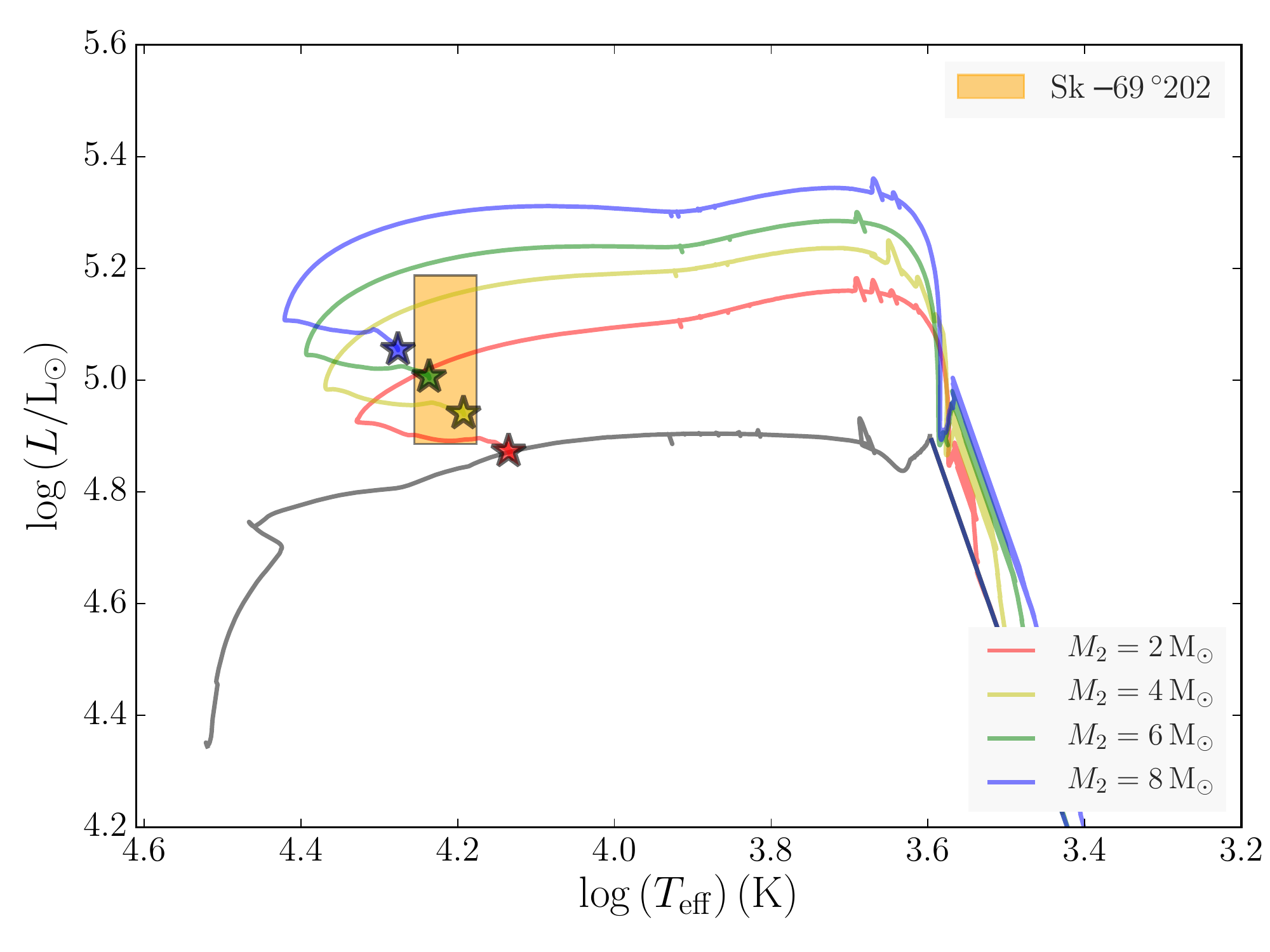}
\caption{HRD of four tracks computed for the merger of $M_{1}=16\,\text{M}_{\odot}$ with $M_{2}=2\,\text{M}_{\odot}$, $4\,\text{M}_{\odot}$, $6\,\text{M}_{\odot}$, and $8\,\text{M}_{\odot}$, and $f_\text{sh}=50\,\%$.  Stars denote the pre-SN models of individual evolutionary tracks. The shaded orange region denotes the observation limits as in Fig.~\ref{fcore}.  \label{hrd_m2}}
\end{figure*}

\begin{figure*}
\centering\includegraphics[width=0.95\textwidth]{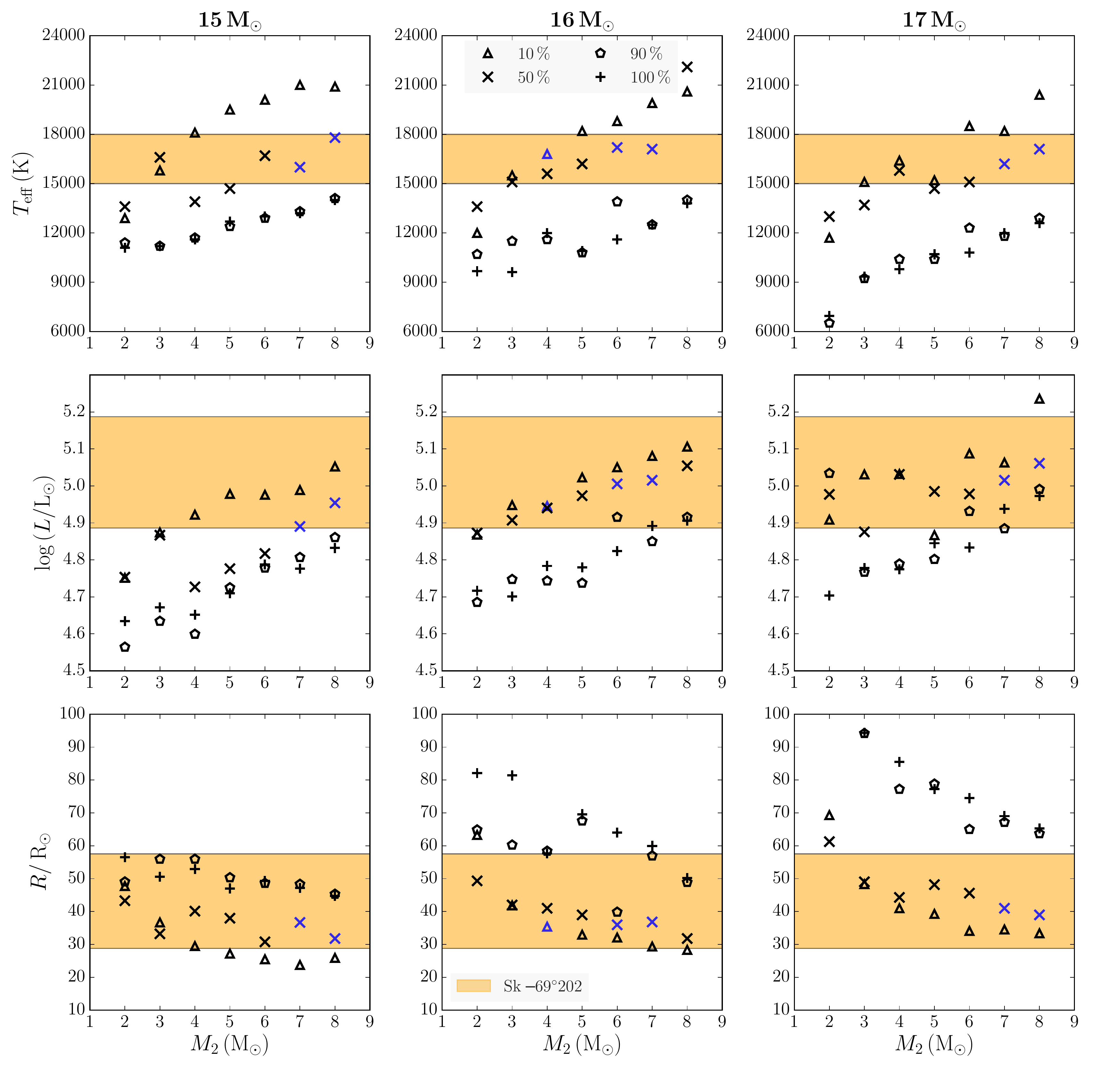}
\caption{Distribution of effective temperature ($T_\text{eff}$), luminosity ($L$) and radius ($R$) of all pre-SN models computed in this work.  Symbols are as in Fig.~\ref{nc-no-m2}. The shaded orange region denotes the observation limits as in Fig.~\ref{fcore}. \label{l-teff-m2}}
\end{figure*}

\begin{figure*}
\centering\includegraphics[width=0.7\textwidth]{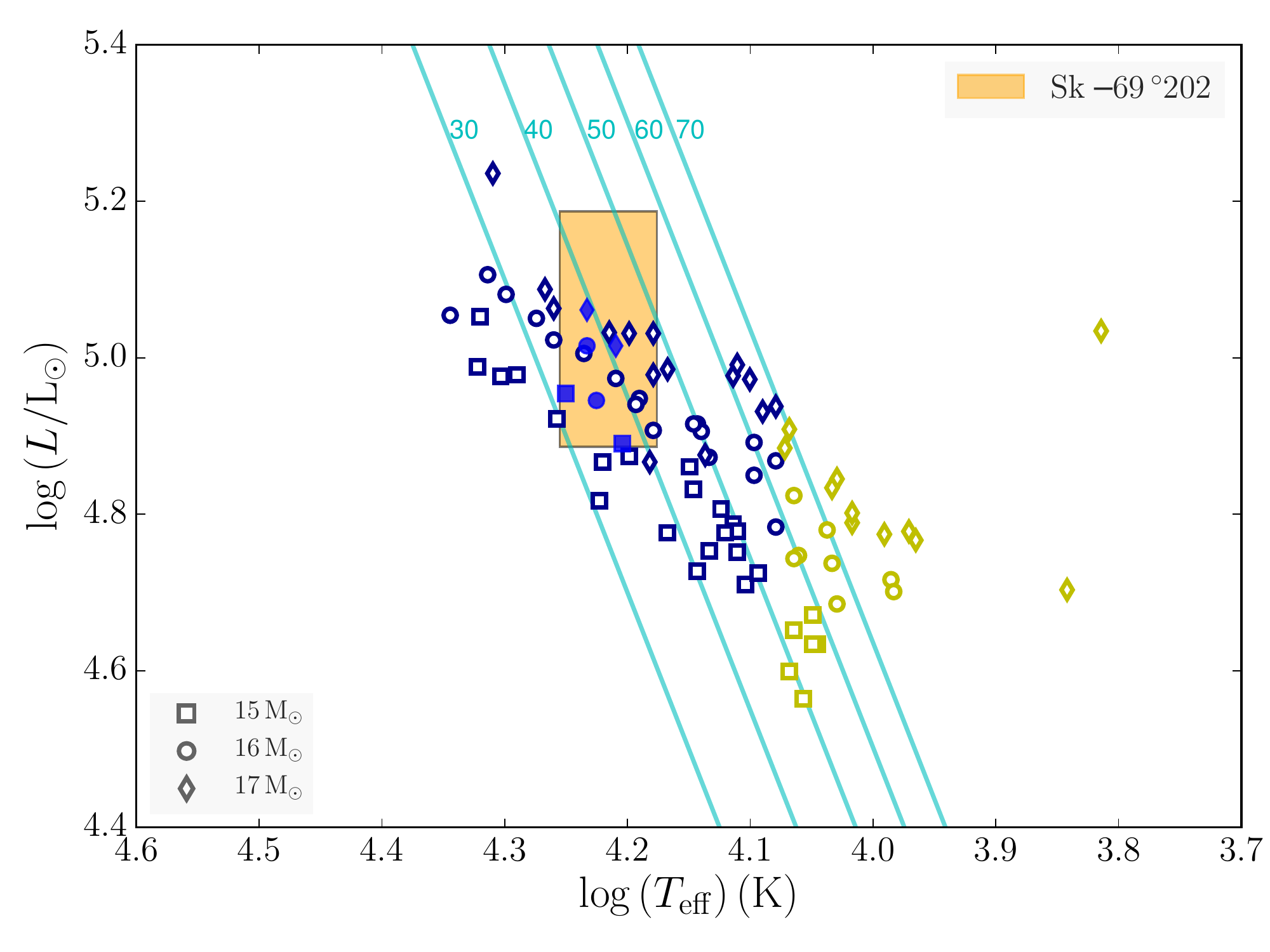}
\caption{Distribution of all 84 pre-SN models in the HRD. Lines of constant surface radius (in $\text{R}_{\odot}$) are drawn.
The shaded orange region denotes the observation limits as in Fig.~\ref{fcore}. Symbols are as explained in Fig.~\ref{nc_no}. \label{hrd_rad}}
\end{figure*}

Varying $f_\text{c}$ affects the effective temperature, $T_\text{eff}$, and luminosity, $L$, of the merged star (Fig.~\ref{fcore}). Increasing $f_\text{c}$ (or $f_\text{sh})$ decreases the He core mass and causes $T_\text{eff}$ and $L$ to also decrease.  In particular for these models, dredging up more than $15\,\%$ of $M_\text{He\,c,1}$ ($50\,\%$ of the He-shell mass), brings down $T_\text{eff}$ from $18\,\text{kK}$ to nearly $12\,\text{kK}$ and it remains more or less constant for larger values of $f_\text{c}$. 

It is interesting to note the case where $m_\text{b}$ is set on the boundary of the CO core (i.e., $f_\text{c}=0$), also becomes a BSG ($T_\text{eff}\approx20\,\text{kK}$). In the two models where $m_\text{b}$ is set above the He core, the post-merger star explodes as a cool RSG of $T_\text{eff}=3\,\text{kK}$.

By increasing $f_\text{sh}$ (Fig.~\ref{hrd_f}), the secondary star mixes deeper inside the He core and the stars become brighter and hotter after the merger but evolve further from the bluest point of their evolution to the cooler and less luminous regions of the HRD. Thus reducing the He core mass for a given $M_{1}$ and $M_{2}$ causes the pre-SN model to become redder. On the other hand for a fixed value of $f_\text{sh}$,  which fixes the He core mass of the post-merger star,  increasing $M_{2}$ increases the envelope mass $M_\text{env}$ of the post-merger star and makes it hotter and more luminous throughout its evolution (Fig.~\ref{hrd_m2}). 

We arrive at two results at this point--  in order to obtain high values of N/C and N/O at the surface and high values of  $T_\text{eff}$ and $L$ required for the progenitor of SN~1987A, we need to restrict the mixing boundary $m_\text{b}$, to be inside the He shell or on the He core boundary.  Second, without He core penetration during the merger, the post-merger stars evolve as cool RSGs until they explode. Let us now understand how varying  $f_\text{sh}$ and $M_{2}$  for a particular $M_{1}$ affect the evolution of the post-merger star.

From Fig.~\ref{l-teff-m2}, we see that for a fixed combination of $M_{1}$ and $M_{2}$, increasing $f_\text{sh}$ makes the pre-SN model cooler. On the other hand for a fixed value of $f_\text{sh}$, increasing $M_{2}$ increases the envelope mass and makes the final model hotter and also more luminous. Since radius drops as a power of $T^{4}$ (for a fixed luminosity), with increasing temperatures the radius of the pre-SN model reduces. The frequency of YSG   pre-SN stars increase as  $M_{1}$ increases, with $f_\text{sh}=90\,\%$ and $100\,\%$ and for smaller values of $M_{2}$.

We now arrive at our next set of conclusions. A merged star is most likely to end its life as a BSG and have high values of N/C and N/O in the surface across all values of $M_{1}$ and $M_{2}$ used in this study, if $f_\text{sh}$ is between $10\,\%-50\%$. The frequency of cooler stars ( $T_\text{eff}<12$\,kK) increases as $M_{1}$ and $f_\text{sh}$ increase and $M_{2}$ decreases. We observe that the most important initial parameter that affects $T_\text{eff}$ of the pre-SN model is $f_\text{sh}$ which determines the He core mass, followed by $M_{2}$ which determines the envelope mass and finally, $M_{1}$. This suggests that there must exist an underlying connection between $T_\text{eff}$ and core-to-envelope mass ratio of the post-merger star.

\subsubsection{Lifetime of BSG model before explosion}
We address the final quantity that has been observationally deduced-- the duration of the BSG phase of the progenitor, $\tau_\text{BSG}$. We calculate $\tau_\text{BSG}$ as the period when the post-merger star attains $T_\text{eff}=12$\,kK until the time of its explosion. From Table~\ref{presn}, our BSG progenitors for SN~1987A have lifetimes that are larger than $15\,\text{kyr}-20\,\text{kyr}$ expected from observations (and is the case for all the BSG pre-SN models obtained in this study, see Section~\ref{appendix}). This parameter however, does not depend on the three initial parameters we varied, but in fact on the age of the primary RSG model just before the merger begins. The younger the RSG model (the earlier along the giant branch it is), the longer the post-merger remnant lives as a BSG. If it is too late along the giant branch, the core is too close to carbon-ignition and $\tau_\text{BSG}$ decreases. A proxy for the age of the RSG model is the mass fraction of helium at the centre ($X_\text{He\,c,1}$), which decreases as the RSG model grows older. The RSG primary models in this study were chosen when $X_\text{He\,c,1}\sim10^{-2}$. Table~\ref{hedep} lists the values of various parameters of the pre-SN models from the merger of different RSG models of $M_{1}=16\,\text{M}_{\odot}$ and a fixed secondary mass, $M_{2}=6\,\text{M}_{\odot}$, and $f_\text{sh}=50\,\%$. As $X_\text{He\,c,1}$ decreases, $\tau_\text{BSG}$ of the post-merger model also decreases. The other parameters of the pre-SN model are largely unaffected.

\begin{table*}
 \caption{Properties of different RSG models of $M_{1}=16\,\text{M}_{\odot}$, $\omega/\omega_\text{crit}=0.30$ and pre-SN models from their merger with $M_{2}=6\,\text{M}_{\odot}$ and $f_\text{sh}=50\,\%$. $X_\text{He\,c}$, $\rho_\text{c}$, $T_\text{c}$, $R$, and $M_\text{He\,c,1}$ are central helium mass fraction, central density, central temperature, radius and He core mass of RSG model; log\,$(L)$, $T_\text{eff}$, $R_\text{pre-SN}$, $T_\text{eff}$ are the luminosity, effective temperature and radius of the pre-SN model; N/C, N/O, He/H are the number ratios of nitrogen to carbon, nitrogen to oxygen and helium to hydrogen of the pre-SN model;  $\tau_\text{BSG}$ is the lifetime of BSG before explosion.}
 \label{hedep}
 \begin{tabular}{lccccccccccc}
  \hline
$X_\text{He\,c}$ & $\rho_\text{c}$ & $T_\text{c}$ & $R_\text{RSG}$ & $M_\text{He\,c,1}$ & log\,$(L)$ & $T_\text{eff}$& $R_\text{pre-SN}$ & N/C & N/O & He/H & $\tau_\text{BSG}$\\

& ($10^{3}$\,g/cc) & ($10^{8}$\,K) & ($\text{R}_{\odot})$ & ($\text{M}_\odot$) & ($\text{L}_{\odot}$) & (kK) & ($\text{R}_{\odot})$ & & & & (kyr)\\
  \hline
$10^{-2}$ & 2.5 & 2.6 & 607 & 4.92 & 4.97 & 16.7 & 43.1 & 8.1 & 1.5 & 0.15 & 48.0 \\
$10^{-4}$ & 4.4 & 3.1 & 773 & 4.94 & 4.94 & 16.6 & 36.5 & 7.6 & 1.42 & 0.14 & 18.3 \\
$10^{-6}$ & 5.5 & 3.3 & 778 & 4.94 & 5.07 & 16.7 & 41.4 & 8.0 & 1.43 & 0.14 & 17.2 \\
$10^{-8}$ & 5.5 & 3.4 & 824 & 4.94 & 5.02 & 16.3 & 41.4 & 8.2 & 1.43 & 0.14 & 17.1 \\
  \hline
 \end{tabular}
\end{table*}

We have thus demonstrated that it is possible to obtain BSG progenitors for Type II SNe, with a range of luminosities, effective temperatures, envelope compositions and durations of the BSG phase  from various combinations of initial parameters for binary mergers. We summarize our results in Section~\ref{conclusions}.

\section{Discussions and Conclusions}
\label{conclusions}

In this paper, we present Type II SN progenitors from the first detailed stellar evolution study of binary mergers of massive stars. We provide details of the 84 pre-SN models computed in Section~\ref{appendix}. Our specific goal was to replicate the observational signatures of the BSG progenitor of SN~1987A, Sk --$69\,^{\circ}202$. For this purpose, we evolved merger models until the pre-SN stage, from a parameter space consisting of the primary mass $M_{1}=15\,\text{M}_{\odot}-17\,\text{M}_{\odot}$, the secondary mass $M_{2}=2\,\text{M}_{\odot}-8\,\text{M}_{\odot}$ and the fraction of He shell dredged up from the He core, $f_\text{sh}=10\,\%,\;50\,\%,\;90\,\%,\;100\,\%$.  Within the evolutionary scenario and parameter space explored, we find that Sk --$69\,^{\circ}202$ can be reproduced with different combinations of the above three parameters. The nature of the pre-SN models rely only on the choice of these three parameters and no additional fine-tuning is required during the evolution of the star to produce BSGs. 
All the pre-SN models have $T_\mathrm{eff}\ge7$\,K, majority of which are BSGs ($T_\mathrm{eff}\ge12$\,K). This leads us to conclude that the progenitors of Type II-pec SNe are highly favoured outcomes from a binary merger.

We draw the following inferences from our results:
\begin{enumerate}[1.]
\item The parameter of paramount importance that determines how hot the surface of the pre-SN model is, is $f_\text{sh}$.  Across the range of primary and secondary masses chosen, BSG pre-SN models are produced when the boundary of mixing is set within $f_\text{sh}=10\,\%-50\,\%$.  These values of $f_\text{sh}$ result in high values of N/C and N/O in the surface as well.

\item The second parameter of importance is the mass of the secondary star, $M_{2}$.  Increasing $M_{2}$ for a fixed value of $f_\text{sh}$ (which determines the post-merger He core mass), increases the $T_\text{eff}$ of the pre-SN star but decreases N/C and N/O in the surface.

\item Finally, the parameter that affects the lifetime of the BSG star ($\tau_\text{BSG}$) before its explosion, is the age of the primary RSG model at the time of the merger. Younger RSG models lead to longer-lived post-merger BSG models.

\item YSG progenitors are produced when either $M_{2}=2\Msun$ OR for small He cores, i.e. when  $f_\text{sh}>50\,\%$. These models increase in number as $M_{1}$ increases. The only conditions under which RSG progenitors are produced, is when the mixing boundary  is set above the He core, i.e., the He core is not penetrated in the merger.

\item The pre-SN models that match Sk --$69\,^{\circ}202$ are from the following systems: 
$M_{1}+M_{2}\,(f_\text{is})$=
$15+7\,\text{M}_{\odot}\,(50\,\%)$, 
$15+8\,\text{M}_{\odot}\,(50\,\%)$, $16+4\,\text{M}_{\odot}\,(10\,\%)$, $16+7\,\text{M}_{\odot}\,(50\,\%)$, $17+7\,\text{M}_{\odot}\,(50\,\%)$, $17+8\,\text{M}_{\odot}\,(10\,\%)$.

\item BSGs are found to span a large range of  N/C and N/O values in the surface ($\mathrm{N/C}=1.8-13$, $\mathrm{N/O}=0.4-1.8$) whereas YSGs are found almost entirely within $\mathrm{N/C}=1-9.7$ and $\mathrm{N/O}=0.4-1.0$. He/H in the surface is between $0.13-0.17$ in all pre-SN models. 

\end{enumerate}

Our results indicate a correlation between $T_\text{eff}$ of a pre-SN model and its core-to-envelope mass ratio along with the fractional decrease of the He core after the merger. The role of small He cores and large envelopes in making blue stars have been discussed by \citet{barkat1989}, \citet{pods1992a}, \citet{woosley1997}, \citet{vanbeveren2013} and \citet{petermann2015}.

Despite our best efforts, we were unable to draw a relation between the three initial parameters, $M_{1}, M_{2}, f_\mathrm{sh}$. In this context we would like to draw parallels with the results of \citet{pods1992a}. In their study, which has a similar initial merger system as ours, they found that as 
the He core to total mass ratio (q) decreases, the pre-SN radius (or $T_\text{eff}$) decreases monotonically (Fig. 11 in \citealt{pods1992a}). They even found a critical value for q from their models-- those post-merger stars that have $q<0.27$ become BSGs. Simply put, larger envelope masses (by accreting larger secondary masses) relative to the He core mass in the post-merger star can result in BSGs. As a corollary, smaller core masses for a given envelope mass should lead to BSGs as well, which we found not to be true in our study. Although there is a general upward trend in $T_\text{eff}$ as $M_{2}$ increases, decreasing the He core mass beyond $50\%$ of the He shell, can cause $T_\text{eff}$ to reduce for the same $M_{2}$.

There may be other reasons as to why BSGs form-- \citet{ivanovathesis,vanbeveren2013} mention that the sharp rise in the hydrogen profile between the He core and the envelope after the merger or the additional fuel supplied to the H-burning shell may also be causes. We hope that our findings will contribute to the quest of understanding why stars end up becoming becoming BSGs or for that matter, RSGs.

Our pre-SN models span a large range of N/C and N/O ratios at the surface, demonstrating that chemical abundances and the position in the HRD of the progenitor are independent constraints.  Simultaneously obtaining the right surface abundance ratios and the HRD position of Sk --$69\,^{\circ}202$ has not been achievable by any single-star model.  Using higher-mass mergers, we can obtain larger N/O ratios and luminosity, comparable to what is found for Sher 25, which has $\log\left(L/\text{L}_{\odot}\right)>5.78-5.90$ \citep{smartt2002,melena2008} and $\mathrm{N/O}\sim1.7-2.1$ \citep{hendry2008}. In the same vein, we can compare our models with the circumstellar abundances and HRD positions of other BSGs that have ring nebulae around them to confirm their origin from binary mergers.

The majority of Type II-pec SNe found so far have been in low-metallicity galaxies and hence \citet{pastorello2012,taddia2013} suggest that low metallicities may play a role in forming BSG progenitors. There maybe an influence of metallicity on the interactions of binary systems-- \citet{demink2008} found that case C mass transfers from massive stars are more likely in low-metallicity environments than in those of solar metallicity. \citet{Egg02} find that the red-to-blue supergiant ratio decreases with metallicity. In our work, the key factor which determines the fate of the pre-SN model is the core-envelope mass ratio of the post-merger star. In order to pursue the question of how likely these mergers are in low-metallicity environments, we need to perform a population synthesis study.

The abundances of Ba and Sr in the surface of our pre-SN models are unchanged from their initial amounts and hence do not exhibit the s-process overabundance detected by \citet{mazzali1992, mazzali1995}. More recent studies, such as those by \citet{utrobin2005} and \citet{dessart2008}, have shown the importance of time-dependent hydrogen ionization in the evolution of Type II SN spectrum. From the time-dependent ionization models for SN~1987A, \citet{utrobin2005} concluded that the barium abundance in its atmosphere matched the LMC value and was not in fact enhanced.

The N/C and N/O ratios in the surface do not vary much from the end of the merger to core collapse. This may suggest that the outer rings likely formed from material ejected by the wind after the merger, but we cannot provide a more detailed dating based on abundance patterns.

The lifetime of the post-merger models as BSGs is $\tau_\text{BSG}=38\,\text{kyr}-149\,\text{kyr}$, which is higher than the $15\,\text{kyr}-20\,\text{kyr}$ estimated for Sk --$69\,^{\circ}202$. We can obtain shorter BSG lifetimes by using older RSG models, keeping all other initial parameters constant.

We do not include the spin-up of the common envelope, or the heating of accreted material in our model, which we intend to look into as part of future work. These effects may affect the evolutionary path of the stars and also help gauge how fast the core will be rotating at the time of explosion.

Mass ejection from the CE phase is not explicitly modelled in this study.  Since no circumstellar disk has been found around the remnant, we assume that the material ejected from the CE is in the nebula alone. The effect of  mass loss from the merger is to cause the envelope mass to reduce and thereby increase the core-to-envelope mass ratio of the post-merger model.  As we accrete a wide range of secondary masses, $2\,\text{M}_{\odot}-8\,\text{M}_{\odot}$, we indirectly explore how mass ejection will affect the structure of the final star. Within the range of secondary masses and the age of the primary model at the time of merger, helium is enhanced by a maximum of $9\,\%$ in the post-merger envelope compared to its initial value, which did not leave a significant impact on the final state of our stars.  We thus rule out the role of helium in obtaining BSGs from mergers. 

Overall, our BSG pre-SN models are more massive than single-star progenitor models for SN~1987A.  With single-star models, the progenitor mass is determined by first comparing the luminosity of the He core of the pre-SN model with the luminosity on the surface \citep{woosley1988b, smartt2009b, dessart2010a}. The reason this can be done for single-star models is that mass loss from the surface has little effect on the He core mass, hence the He core mass is strongly correlated to the ZAMS mass of the star \citep{dessart2010a}. The He core mass thus determined for the luminosity of Sk --$69\,^{\circ}202$ is  $M_\text{He\,c}=4\,\text{M}_{\odot}-7\,\text{M}_{\odot}$, which would originate from a ZAMS star of mass, $M_\text{ZAMS}=14\,\text{M}_{\odot}-20\,\text{M}_{\odot}$ \citep{arnett1989,smartt2009a,smartt2009b}. In the case of our merger models, the pre-SN He core mass depends on $M_{1}$ and $m_\text{b}$ whereas the initial mass is the sum of $M_{1}$ and $M_{2}$ and hence, a given He core mass could belong to any number of initial masses depending on the accreted value of  $M_{2}$. 

Consequently these merger models will impact the parametrised studies of SN explosion properties that are calibrated against supernova SN~1987A.  Typical single-star models used for SN~1987A are those from \citet{woosley1988a} and \citet{woosley1988b},  $M_\text{ZAMS}=15\,\text{M}_{\odot}-20\,\text{M}_{\odot}$, $M_\text{He\,c}=4.1\,\text{M}_{\odot}-6.2\,\text{M}_{\odot}$ and $M_\text{env}=5\,\text{M}_{\odot}-10\,\text{M}_{\odot}$ \citep{arnett1989,dessart2010b,utrobin2015} or the He-enriched models of \citet{nomoto1988} and \citet{saio1988}, $M_\text{ZAMS}=23\,\text{M}_{\odot}$, $M_\text{He\,c}=6\,\text{M}_{\odot}$ and $M_\text{env}=10.3\,\text{M}_{\odot}$ \citep{blinnikov2000,kleiser2011,ugliano2012}. Scaling relations based on these models of SN~1987A have been used to determine the explosion properties of other Type II SNe as well (e.g., see \citealt{kleiser2011}).

Our BSG pre-SN models have lower He core masses, $M_\text{He\,c}=2.4\,\text{M}_{\odot}-4.5\,\text{M}_{\odot}$, and much larger envelope masses $M_\text{env}=12.3\,\text{M}_{\odot}-20.5\,\text{M}_{\odot}$.  It is therefore imperative to determine the explosion properties of SN~1987A with these models. In a subsequent paper, we will present the light curves and spectra from the explosions of these models using a radiative transfer code and compare them to existing observations of Type II-pec SNe, focusing particularly on SN~1987A.

\section{Appendix}
\label{appendix}
This section contains the properties of all the pre-SN models computed in this study.

\begin{table*}
 \centering
 \caption{Parameters of \textbf{BSG} binary merger progenitors of Type II SNe ($T_\text{eff}\geq 12$\,kK).  Column headings are the same as in Table~\ref{presn}. '--' under tHe\,column $M_\text{Fe\,c}$ are for those runs that crashed while at core-silicon burning and thus did not reach the pre-SN model.}
 {\begin{tabular}{lrrrrrrrrrrrrrrr}
    \hline
$M_\text{1}$ & $M_\text{2}$ & $f_\text{sh}$ & $f_\text{c}$ & $m_\text{b}$ & $M_\text{He\,c}$ & $M_\text{env}$ & $M_\text{pre-SN}$ & $T_\text{eff}$ & $\log\left(L\right)$  & $R_\text{pre-SN}$ & $M_\text{Fe\,c}$ & N/C & N/O & He/H & $\tau_\text{BSG}$\\

($\text{M}_{\odot}$) & ($\text{M}_{\odot}$) & $\%$ & $\%$ & ($\text{M}_{\odot}$) & ($\text{M}_{\odot}$) & ($\text{M}_{\odot}$) & ($\text{M}_{\odot}$) & (kK) & ($\text{L}_\odot$) & ($\text{R}_{\odot}$) & ($\text{M}_{\odot}$) & & & & (kyr)\\
  \hline
15 & 2 & 10 &  3.5 & 4.29 & 3.75 & 12.33 & 16.08 & 12.9 & 4.75 & 47.8 & 1.41 & 8.8 & 1.4 & 0.14 & 85 \\ 
15 & 2 & 50 & 17.5 & 3.67 & 3.40 & 12.71 & 16.09 & 13.6 & 4.75 & 43.3 & 1.47 & 11.2& 1.6 & 0.15 & 76 \\ 
15 & 3 & 10 &  3.5 & 4.29 & 3.67 & 13.41 & 17.08 & 15.8 & 4.87 & 36.7 & 1.58 & 7.9 & 1.4 & 0.13 & 65 \\ 
15 & 3 & 50 & 17.5 & 3.67 & 3.40 & 13.68 & 17.08 & 16.6 & 4.87 & 33.2 & 1.44 & 9.0 & 1.5 & 0.15 & 78 \\ 
15 & 4 & 10 &  3.5 & 4.29 & 3.65 & 14.43 & 18.08 & 18.1 & 4.92 & 29.5 & 1.58 & 7.0 & 1.3 & 0.13 & 66 \\ 
15 & 4 & 50 & 17.5 & 3.67 & 2.85 & 15.22 & 18.07 & 13.9 & 4.73 & 40.1 & -- & 8.2 & 1.4 & 0.14 & 101 \\ 
15 & 5 & 10 &  3.5 & 4.29 & 3.51 & 15.56 & 19.07 & 19.5 & 4.98 & 27.2 & 1.52 & 6.4 & 1.3 & 0.13 & 71 \\ 
15 & 5 & 50 & 17.5 & 3.67 & 2.86 & 16.21 & 19.07 & 14.7 & 4.78 & 38.0 & 1.46 & 7.5 & 1.4 & 0.14 & 105 \\ 
15 & 5 & 100& 35.0 & 2.89 & 2.35 & 16.70 & 19.05 & 12.7 & 4.71 & 47.1 & -- & 2.1 & 0.5 & 0.15 & 149 \\ 
15 & 5 & 90 & 31.5 & 3.05 & 2.59 & 16.47 & 19.06 & 12.4 & 4.72 & 50.4 & -- & 5.6 & 0.9 & 0.15 & 104 \\ 
15 & 6 & 10 &  3.5 & 4.29 & 3.57 & 16.50 & 20.07 & 20.1 & 4.98 & 25.5 & 1.54 & 5.9 & 1.2 & 0.12 & 71 \\ 
15 & 6 & 50 & 17.5 & 3.67 & 3.00 & 17.07 & 20.07 & 16.7 & 4.82 & 30.8 & 1.49 & 6.8 & 1.3 & 0.14 & 77 \\ 
15 & 6 & 100& 35.0 & 2.89 & 2.65 & 17.40 & 20.05 & 13.0 & 4.79 & 49.4 & -- & 4.2 & 0.8 & 0.15 & 95 \\ 
15 & 6 & 90 & 31.5 & 3.05 & 2.44 & 17.61 & 20.05 & 12.9 & 4.78 & 48.6 & -- & 2.4 & 0.5 & 0.15 & 133 \\ 
15 & 7 & 10 &  3.5 & 4.29 & 3.57 & 17.49 & 21.06 & 21.0 & 4.99 & 23.7 & 1.49 & 5.5 & 1.2 & 0.07 & 82 \\ 
15 & 7 & 50 & 17.5 & 3.67 & 2.90 & 18.16 & 21.06 & 16.0 & 4.89 & 36.7 & 1.46 & 6.5 & 1.3 & 0.13 & 82 \\ 
15 & 7 & 100& 35.0 & 2.89 & 2.40 & 18.63 & 21.03 & 13.2 & 4.78 & 47.2 & -- & 1.5 & 0.4 & 0.15 & 132 \\ 
15 & 7 & 90 & 31.5 & 3.05 & 2.50 & 18.55 & 21.05 & 13.3 & 4.81 & 48.3 & -- & 4.3 & 0.8 & 0.14 & 126 \\ 
15 & 8 & 10 &  3.5 & 4.29 & 3.32 & 18.73 & 22.05 & 20.9 & 5.05 & 25.9 & 1.56 & 5.2 & 1.2 & 0.12 & 69 \\ 
15 & 8 & 50 & 17.5 & 3.67 & 2.95 & 19.10 & 22.05 & 17.8 & 4.95 & 31.8 & 1.39 & 5.8 & 1.3 & 0.13 & 83 \\ 
15 & 8 & 100& 35.0 & 2.89 & 2.35 & 19.68 & 22.03 & 14.0 & 4.83 & 44.7 & -- & 2.5 & 0.6 & 0.14 & 133 \\ 
15 & 8 & 90 & 31.5 & 3.05 & 2.56 & 19.48 & 22.04 & 14.1 & 4.86 & 45.3 & -- & 3.4 & 0.7 & 0.14 & 102 \\ 
16 & 2 & 10 &  3.3 & 4.71 & 4.16 & 12.85 & 17.01 & 12.0 & 4.87 & 63.3 & 1.64 & 9.2 & 1.5 & 0.14 & 38 \\ 
16 & 2 & 50 & 16.6 & 4.06 & 3.73 & 13.28 & 17.01 & 13.6 & 4.87 & 49.4 & -- & 11.6& 1.6 & 0.16 & 62 \\ 
16 & 3 & 10 &  3.3 & 4.71 & 4.15 & 13.85 & 18.00 & 15.5 & 4.95 & 41.9 & 1.68 & 8.1 & 1.5 & 0.14 & 66 \\ 
16 & 3 & 50 & 16.6 & 4.06 & 3.60 & 14.40 & 18.00 & 15.1 & 4.91 & 42.0 & -- & 10.0& 1.6 & 0.15 & 68 \\ 
16 & 4 & 10 &  3.3 & 4.71 & 4.11 & 14.89 & 19.00 & 16.8 & 4.95 & 35.4 & 1.65 & 6.6 & 1.4 & 0.13 & 41 \\ 
16 & 4 & 50 & 16.6 & 4.06 & 3.63 & 15.37 & 19.00 & 15.6 & 4.94 & 41.0 & 1.58 & 9.0 & 1.5 & 0.15 & 48 \\ 
16 & 4 & 100& 33.2 & 3.25 & 2.86 & 16.12 & 18.98 & 12.0 & 4.78 & 57.7 & -- & 2.6 & 0.5 & 0.17 & 106 \\ 
16 & 5 & 10 &  3.3 & 4.71 & 4.05 & 15.94 & 19.99 & 18.2 & 5.02 & 32.9 & 1.68 & 6.5 & 1.3 & 0.13 & 51 \\ 
16 & 5 & 50 & 16.6 & 4.06 & 3.65 & 16.34 & 19.99 & 16.2 & 4.97 & 39.0 & -- & 8.1 & 1.5 & 0.15 & 47 \\ 
16 & 6 & 10 &  3.3 & 4.71 & 3.97 & 17.02 & 20.99 & 18.8 & 5.05 & 32.1 & -- & 6.1 & 1.3 & 0.13 & 55 \\ 
16 & 6 & 50 & 16.6 & 4.06 & 3.61 & 17.38 & 20.99 & 17.2 & 5.01 & 36.0 & 1.57 & 7.2 & 1.4 & 0.14 & 48 \\ 
16 & 6 & 90 & 30.0 & 3.41 & 3.09 & 17.89 & 20.98 & 13.9 & 4.92 & 39.9 & 1.52 & 8.3 & 1.1 & 0.15 & 74 \\ 
16 & 7 & 10 &  3.3 & 4.71 & 3.85 & 18.14 & 21.99 & 19.9 & 5.08 & 29.4 & -- & 5.6 & 1.3 & 0.13 & 41 \\ 
16 & 7 & 50 & 16.6 & 4.06 & 3.41 & 18.57 & 21.98 & 17.1 & 5.02 & 36.8 & -- & 6.9 & 1.4 & 0.14 & 49 \\ 
16 & 7 & 100& 33.2 & 3.25 & 2.98 & 18.99 & 21.97 & 12.5 & 4.89 & 60.0 & 1.47 & 3.0 & 0.7 & 0.15 & 63 \\ 
16 & 7 & 90 & 30.0 & 3.41 & 2.82 & 19.15 & 21.97 & 12.5 & 4.85 & 57.0 & 1.44 & 7.0 & 1.2 & 0.15 & 67 \\ 
16 & 8 & 10 &  3.3 & 4.71 & 3.84 & 19.14 & 22.98 & 20.6 & 5.11 & 28.3 & -- & 5.3 & 1.2 & 0.13 & 42 \\ 
16 & 8 & 50 & 16.6 & 4.06 & 3.52 & 19.45 & 22.97 & 22.1 & 5.05 & 31.8 & -- & 6.0 & 1.3 & 0.13 & 60 \\ 
16 & 8 & 100& 33.2 & 3.25 & 2.91 & 20.04 & 22.95 & 13.8 & 4.91 & 50.2 & 1.44 & 2.6 & 0.6 & 0.15 & 87 \\ 
16 & 8 & 90 & 30.0 & 3.41 & 2.93 & 20.03 & 22.96 & 14.0 & 4.92 & 48.9 & 1.60 & 5.4 & 0.8 & 0.15 & 89 \\ 
17 & 2 & 50 & 15.6 & 4.44 & 4.07 & 13.77 & 17.84 & 13.0 & 4.98 & 61.3 & 1.61 & 12.5& 1.7 & 0.17 & 40 \\ 
17 & 3 & 10 &  3.1 & 5.10 & 4.62 & 14.22 & 18.84 & 15.1 & 5.03 & 48.3 & 1.68 & 8.4 & 1.5 & 0.14 & 34 \\ 
17 & 3 & 50 & 15.6 & 4.44 & 4.24 & 14.59 & 18.84 & 13.7 & 4.88 & 49.1 & 1.62 & 10.1& 1.6 & 0.16 & 39 \\ 
17 & 4 & 10 &  3.1 & 5.10 & 4.49 & 15.35 & 19.84 & 16.4 & 5.03 & 41.0 & -- & 12.7& 1.4 & 0.14 & 36 \\ 
17 & 4 & 50 & 15.6 & 4.44 & 4.30 & 15.53 & 19.83 & 15.8 & 5.03 & 44.3 & 1.68 & 8.6 & 1.5 & 0.15 & 37 \\ 
17 & 5 & 10 &  3.1 & 5.10 & 4.48 & 16.35 & 20.83 & 15.2 & 4.87 & 39.3 & 1.79 & 6.8 & 1.4 & 0.14 & 39 \\ 
17 & 5 & 50 & 15.6 & 4.44 & 3.93 & 16.90 & 20.83 & 14.7 & 4.99 & 48.2 & 1.65 & 8.5 & 1.5 & 0.15 & 41 \\ 
17 & 6 & 10 &  3.1 & 5.10 & 4.34 & 17.48 & 21.82 & 18.5 & 5.09 & 34.1 & -- & 6.2 & 1.3 & 0.13 & 35 \\ 
17 & 6 & 50 & 15.6 & 4.44 & 3.88 & 17.94 & 21.82 & 15.1 & 4.98 & 45.6 & 1.67 & 7.8 & 1.5 & 0.14 & 41 \\ 
17 & 6 & 90 & 28.1 & 3.78 & 3.29 & 18.51 & 21.80 & 12.3 & 4.93 & 65.0 & 1.59 & 1.9 & 0.5 & 0.16 & 57 \\ 
17 & 7 & 10 &  3.1 & 5.10 & 4.25 & 18.57 & 22.82 & 18.2 & 5.06 & 34.5 & 1.65 & 5.9 & 1.3 & 0.13 & 39 \\ 
17 & 7 & 50 & 15.6 & 4.44 & 3.86 & 18.95 & 22.81 & 16.2 & 5.02 & 41.0 & 1.66 & 7.0 & 1.4 & 0.14 & 41 \\ 
17 & 7 & 100& 31.2 & 3.62 & 3.21 & 19.60 & 22.79 & 12.0 & 4.94 & 69.1 & 1.65 & 1.6 & 0.4 & 0.16 & 54 \\ 
17 & 8 & 10 &  3.1 & 5.10 & 4.24 & 19.57 & 23.81 & 20.4 & 5.24 & 33.4 & 1.81 & 5.5 & 1.3 & 0.13 & 40 \\ 
17 & 8 & 50 & 15.6 & 4.44 & 3.83 & 19.98 & 23.81 & 17.1 & 5.06 & 39.0 & 1.71 & 6.4 & 1.4 & 0.14 & 41 \\ 
17 & 8 & 100& 31.2 & 3.62 & 3.32 & 20.47 & 23.79 & 12.6 & 4.97 & 65.3 & 1.62 & 4.0 & 0.8 & 0.15 & 50 \\ 
17 & 8 & 90 & 28.1 & 3.78 & 3.35 & 20.44 & 23.79 & 12.9 & 4.99 & 63.7 & 1.67 & 4.9 & 0.9 & 0.15 & 54 \\ 

  \hline
 \end{tabular}}
 \label{appendix_table}
\end{table*}

\begin{table*}
 \centering
 \caption{Parameters of \textbf{YSG} binary merger progenitors of Type II SNe  ($4\,\text{kK}\leq T_\text{eff}< 12\,\text{kK}$). Column headings are same as Table~\ref{appendix_table}.}
 {\begin{tabular}{lrrrrrrrrrrrrrrr}
  \hline
$M_\text{1}$ & $M_\text{2}$ & $f_\text{sh}$ & $f_\text{c}$ & $m_\text{b}$ & $M_\text{He\,c}$ & $M_\text{env}$ & $M_\text{pre-SN}$ & $T_\text{eff}$ & $\log\left(L\right)$  & $R_\text{pre-SN}$ & $M_\text{Fe\,c}$ & N/C & N/O & He/H & $\tau_\text{BSG}$\\

($\text{M}_{\odot}$) & ($\text{M}_{\odot}$) & $\%$ & $\%$ & ($\text{M}_{\odot}$) & ($\text{M}_{\odot}$) & ($\text{M}_{\odot}$) & ($\text{M}_{\odot}$) & (kK) & ($\text{L}_\odot$) & ($\text{R}_{\odot}$) & ($\text{M}_{\odot}$) & & & & (kyr)\\
  \hline
15 & 2 & 100 & 35.0 & 2.89 & 2.68 & 13.40 & 16.08 & 11.1 & 4.63 & 56.5 & 1.41 & 4.4 & 0.8 & 0.17 & 0.0 \\ 
15 & 2 &  90 & 31.5 & 3.05 & 2.67 & 14.13 & 16.80 & 11.4 & 4.56 & 49.1 & 1.70 & 5.9 & 0.8 & 0.17 & 0.0 \\ 
15 & 3 & 100 & 35.0 & 2.89 & 2.51 & 14.55 & 17.06 & 11.2 & 4.67 & 50.6 & -- & 1.5 & 0.4 & 0.17 & 0.0 \\ 
15 & 3 &  90 & 31.5 & 3.05 & 2.67 & 14.40 & 17.07 & 11.2 & 4.63 & 56.0 & 1.42 & 2.6 & 0.6 & 0.17 & 0.0 \\ 
15 & 4 & 100 & 35.0 & 2.89 & 2.46 & 15.60 & 18.06 & 11.6 & 4.65 & 52.9 & 1.33 & 1.7 & 0.4 & 0.16 & 0.0 \\ 
15 & 4 &  90 & 31.5 & 3.05 & 2.74 & 15.33 & 18.07 & 11.7 & 4.60 & 56.0 & 1.44 & 9.5 & 1.2 & 0.16 & 0.0 \\ 
16 & 2 & 100 & 33.2 & 3.25 & 3.14 & 13.86 & 17.00 & 9.67 & 4.72 & 82.2 & -- &20.4 & 0.5 & 0.18 & 0.0 \\ 
16 & 2 &  90 & 30.0 & 3.41 & 3.05 & 13.95 & 17.00 & 10.7 & 4.69 & 64.9 & 1.57 & 1.8 & 0.4 & 0.18 & 0.0 \\ 
16 & 3 & 100 & 33.2 & 3.25 & 3.05 & 14.94 & 17.99 & 9.62 & 4.70 & 81.4 & 1.49 & 2.2 & 0.5 & 0.17 & 0.0 \\ 
16 & 3 &  90 & 30.0 & 3.41 & 3.10 & 14.90 & 18.00 & 11.5 & 4.75 & 60.3 & 1.47 & 6.6 & 1.0 & 0.17 & 0.0 \\ 
16 & 4 &  90 & 30.0 & 3.41 & 3.04 & 15.96 & 19.00 & 11.6 & 4.74 & 58.4 & 1.47 & 8.6 & 1.0 & 0.16 & 0.0 \\ 
16 & 5 & 100 & 33.2 & 3.25 & 3.06 & 16.93 & 19.99 & 10.9 & 4.78 & 69.6 & -- & 6.3 & 1.0 & 0.16 & 0.0 \\ 
16 & 5 &  90 & 30.0 & 3.41 & 3.00 & 16.98 & 19.98 & 10.8 & 4.74 & 67.6 & 1.49 & 2.8 & 0.6 & 0.16 & 0.0 \\ 
16 & 6 & 100 & 33.2 & 3.25 & 3.06 & 17.92 & 20.98 & 11.6 & 4.82 & 64.0 & 1.49 & 4.4 & 0.8 & 0.16 & 0.0 \\ 
17 & 2 &  10 &  3.1 & 5.10 & 4.63 & 13.21 & 17.84 & 11.7 & 4.91 & 69.4 & -- & 9.6 & 1.6 & 0.15 & 0.0 \\ 
17 & 2 &  90 & 28.1 & 3.78 & 3.46 & 14.37 & 17.83 & 6.52 & 5.03 & 177.0& 1.67 & 3.0 & 0.6 & 0.19 & 0.0 \\ 
17 & 2 & 100 & 31.2 & 3.62 & 3.44 & 14.39 & 17.83 & 6.95 & 4.70 & 156.8& 1.65 & 2.0 & 0.4 & 0.19 & 0.0 \\ 
17 & 3 & 100 & 31.2 & 3.62 & 3.39 & 15.43 & 18.82 & 9.35 & 4.78 & 94.4 & 1.67 & 1.9 & 0.4 & 0.18 & 0.0 \\ 
17 & 3 &  90 & 28.1 & 3.78 & 3.37 & 15.45 & 18.82 & 9.23 & 4.77 & 94.2 & 1.64 & 1.7 & 0.4 & 0.18 & 0.0 \\ 
17 & 4 & 100 & 31.2 & 3.62 & 3.30 & 16.51 & 19.81 & 9.79 & 4.77 & 85.6 & 1.63 & 1.4 & 0.4 & 0.17 & 0.0 \\ 
17 & 4 &  90 & 28.1 & 3.78 & 3.41 & 16.41 & 19.82 & 10.4 & 4.79 & 77.3 & 1.66 & 6.9 & 1.0 & 0.17 & 0.0 \\ 
17 & 5 & 100 & 31.2 & 3.62 & 3.39 & 17.42 & 20.81 & 10.7 & 4.85 & 77.3 & 1.65 & 4.0 & 0.8 & 0.17 & 0.0 \\ 
17 & 5 &  90 & 28.1 & 3.78 & 3.17 & 17.64 & 20.81 & 10.4 & 4.80 & 78.8 & 1.58 & 1.2 & 0.3 & 0.17 & 0.0 \\ 
17 & 6 & 100 & 31.2 & 3.62 & 3.35 & 18.45 & 21.80 & 10.8 & 4.83 & 74.5 & 1.69 & 2.0 & 0.5 & 0.16 & 0.0 \\ 
17 & 7 &  90 & 28.1 & 3.78 & 3.22 & 19.57 & 22.79 & 11.8 & 4.88 & 67.2 & 1.61 & 1.5 & 0.4 & 0.16 & 0.0 \\ 

  \hline
 \end{tabular}}
\end{table*}

\clearpage
\section*{Acknowledgements}
The authors would like to thank Luc Dessart for his feedback on the manuscript and understanding conclusions for supernova light curves.  We thank Philipp Podsiadlowski for useful discussions about his and Natasha Ivanova's model, on which this paper is based, in particular, and both models and observational data for SN 1987A in general.  We also thank Bernhard M\"uller and John Lattanzio for their feedback and useful comments during the making of this manuscript. We also thank Amanda Karakas, Dorottya Sz\`ecsi and Rebecca Nealon for their feedback on previous versions of the manuscript. We thank Thomas Constantino for helping with the opacity tables in KEPLER and Christian Ritter for providing the initial abundance generator tool. This research was supported by US NSF to JINA-CEE through grant PHY-1430152. AH was supported by an ARC Future Fellowship FT120100363.

\clearpage
\bibliographystyle{mnras}
\bibliography{master} 

\begin{thebibliography}{}
\makeatletter
\relax
\def\mn@urlcharsother{\let\do\@makeother \do\$\do\&\do\#\do\^\do\_\do\%\do\~}
\def\mn@doi{\begingroup\mn@urlcharsother \@ifnextchar [ {\mn@doi@}
  {\mn@doi@[]}}
\def\mn@doi@[#1]#2{\def\@tempa{#1}\ifx\@tempa\@empty \href
  {http://dx.doi.org/#2} {doi:#2}\else \href {http://dx.doi.org/#2} {#1}\fi
  \endgroup}
\def\mn@eprint#1#2{\mn@eprint@#1:#2::\@nil}
\def\mn@eprint@arXiv#1{\href {http://arxiv.org/abs/#1} {{\tt arXiv:#1}}}
\def\mn@eprint@dblp#1{\href {http://dblp.uni-trier.de/rec/bibtex/#1.xml}
  {dblp:#1}}
\def\mn@eprint@#1:#2:#3:#4\@nil{\def\@tempa {#1}\def\@tempb {#2}\def\@tempc
  {#3}\ifx \@tempc \@empty \let \@tempc \@tempb \let \@tempb \@tempa \fi \ifx
  \@tempb \@empty \def\@tempb {arXiv}\fi \@ifundefined
  {mn@eprint@\@tempb}{\@tempb:\@tempc}{\expandafter \expandafter \csname
  mn@eprint@\@tempb\endcsname \expandafter{\@tempc}}}

\bibitem[\protect\citeauthoryear{{Arnett}, {Bahcall}, {Kirshner}  \&
  {Woosley}}{{Arnett} et~al.}{1989}]{arnett1989}
{Arnett} W.~D.,  {Bahcall} J.~N.,  {Kirshner} R.~P.,   {Woosley} S.~E.,  1989,
  \mn@doi [\araa] {10.1146/annurev.aa.27.090189.003213}, \href
  {http://adsabs.harvard.edu/abs/1989ARA\%26..27..629A} {27, 629}

\bibitem[\protect\citeauthoryear{{Asplund}, {Grevesse}, {Sauval}  \&
  {Scott}}{{Asplund} et~al.}{2009}]{asplund2009}
{Asplund} M.,  {Grevesse} N.,  {Sauval} A.~J.,   {Scott} P.,  2009, \mn@doi
  [\araa] {10.1146/annurev.astro.46.060407.145222}, \href
  {http://adsabs.harvard.edu/abs/2009ARA%26A..47..481A} {47, 481}

\bibitem[\protect\citeauthoryear{{Barkat} \& {Wheeler}}{{Barkat} \&
  {Wheeler}}{1989}]{barkat1989}
{Barkat} Z.,  {Wheeler} J.~C.,  1989, \mn@doi [\apj] {10.1086/167650}, \href
  {http://adsabs.harvard.edu/abs/1989ApJ...342..940B} {342, 940}

\bibitem[\protect\citeauthoryear{{Blanco} et~al.,}{{Blanco}
  et~al.}{1987}]{blanco1987}
{Blanco} V.~M.,  et~al., 1987, \mn@doi [\apj] {10.1086/165577}, \href
  {http://adsabs.harvard.edu/abs/1987ApJ...320..589B} {320, 589}

\bibitem[\protect\citeauthoryear{{Blinnikov}, {Lundqvist}, {Bartunov}, {Nomoto}
   \& {Iwamoto}}{{Blinnikov} et~al.}{2000}]{blinnikov2000}
{Blinnikov} S.,  {Lundqvist} P.,  {Bartunov} O.,  {Nomoto} K.,   {Iwamoto} K.,
  2000, \mn@doi [\apj] {10.1086/308588}, \href
  {http://adsabs.harvard.edu/abs/2000ApJ...532.1132B} {532, 1132}

\bibitem[\protect\citeauthoryear{{Brandner}, {Grebel}, {Chu}  \&
  {Weis}}{{Brandner} et~al.}{1997a}]{brandner1997b}
{Brandner} W.,  {Grebel} E.~K.,  {Chu} Y.-H.,   {Weis} K.,  1997a, \mn@doi
  [\apjl] {10.1086/310460}, \href
  {http://adsabs.harvard.edu/abs/1997ApJ...475L..45B} {475, L45}

\bibitem[\protect\citeauthoryear{{Brandner}, {Chu}, {Eisenhauer}, {Grebel}  \&
  {Points}}{{Brandner} et~al.}{1997b}]{brandner1997a}
{Brandner} W.,  {Chu} Y.-H.,  {Eisenhauer} F.,  {Grebel} E.~K.,   {Points}
  S.~D.,  1997b, \mn@doi [\apjl] {10.1086/316795}, \href
  {http://adsabs.harvard.edu/abs/1997ApJ...489L.153B} {489, L153}

\bibitem[\protect\citeauthoryear{{Brott} et~al.,}{{Brott}
  et~al.}{2011}]{brott2011}
{Brott} I.,  et~al., 2011, \mn@doi [\aap] {10.1051/0004-6361/201016114}, \href
  {http://adsabs.harvard.edu/abs/2011A\%26A...530A.116B} {530, A116}

\bibitem[\protect\citeauthoryear{{Burrows} et~al.,}{{Burrows}
  et~al.}{1995}]{burrows1995}
{Burrows} C.~J.,  et~al., 1995, \mn@doi [\apj] {10.1086/176339}, \href
  {http://adsabs.harvard.edu/abs/1995ApJ...452..680B} {452, 680}

\bibitem[\protect\citeauthoryear{{Catchpole} et~al.,}{{Catchpole}
  et~al.}{1988}]{catchpole1988}
{Catchpole} R.~M.,  et~al., 1988, \mn@doi [\mnras] {10.1093/mnras/231.1.75P},
  \href {http://adsabs.harvard.edu/abs/1988MNRAS.231P..75C} {231, 75p}

\bibitem[\protect\citeauthoryear{{Chevalier} \& {Dwarkadas}}{{Chevalier} \&
  {Dwarkadas}}{1995}]{chevalier1995}
{Chevalier} R.~A.,  {Dwarkadas} V.~V.,  1995, \mn@doi [\apjl] {10.1086/309714},
  \href {http://adsabs.harvard.edu/abs/1995ApJ...452L..45C} {452, L45}

\bibitem[\protect\citeauthoryear{{Chita}, {Langer}, {van Marle},
  {Garc{\'{\i}}a-Segura}  \& {Heger}}{{Chita} et~al.}{2008}]{chita2008}
{Chita} S.~M.,  {Langer} N.,  {van Marle} A.~J.,  {Garc{\'{\i}}a-Segura} G.,
  {Heger} A.,  2008, \mn@doi [\aap] {10.1051/0004-6361:200810087}, \href
  {http://adsabs.harvard.edu/abs/2008A\%26A...488L..37C} {488, L37}

\bibitem[\protect\citeauthoryear{{Constantino}, {Campbell}, {Gil-Pons}  \&
  {Lattanzio}}{{Constantino} et~al.}{2014}]{constantino2014}
{Constantino} T.,  {Campbell} S.,  {Gil-Pons} P.,   {Lattanzio} J.,  2014,
  \mn@doi [\apj] {10.1088/0004-637X/784/1/56}, \href
  {http://adsabs.harvard.edu/abs/2014ApJ...784...56C} {784, 56}

\bibitem[\protect\citeauthoryear{{Crotts} \& {Heathcote}}{{Crotts} \&
  {Heathcote}}{2000}]{crotts2000}
{Crotts} A.~P.~S.,  {Heathcote} S.~R.,  2000, \mn@doi [\apj] {10.1086/308141},
  \href {http://adsabs.harvard.edu/abs/2000ApJ...528..426C} {528, 426}

\bibitem[\protect\citeauthoryear{{Crotts} \& {Kunkel}}{{Crotts} \&
  {Kunkel}}{1991}]{crotts1991}
{Crotts} A.~P.~S.,  {Kunkel} W.~E.,  1991, \mn@doi [\apjl] {10.1086/185912},
  \href {http://adsabs.harvard.edu/abs/1991ApJ...366L..73C} {366, L73}

\bibitem[\protect\citeauthoryear{{Dessart} \& {Hillier}}{{Dessart} \&
  {Hillier}}{2008}]{dessart2008}
{Dessart} L.,  {Hillier} D.~J.,  2008, \mn@doi [\mnras]
  {10.1111/j.1365-2966.2007.12538.x}, \href
  {http://adsabs.harvard.edu/abs/2008MNRAS.383...57D} {383, 57}

\bibitem[\protect\citeauthoryear{{Dessart} \& {Hillier}}{{Dessart} \&
  {Hillier}}{2010}]{dessart2010b}
{Dessart} L.,  {Hillier} D.~J.,  2010, \mn@doi [\mnras]
  {10.1111/j.1365-2966.2010.16611.x}, \href
  {http://adsabs.harvard.edu/abs/2010MNRAS.405.2141D} {405, 2141}

\bibitem[\protect\citeauthoryear{{Dessart}, {Livne}  \& {Waldman}}{{Dessart}
  et~al.}{2010}]{dessart2010a}
{Dessart} L.,  {Livne} E.,   {Waldman} R.,  2010, \mn@doi [\mnras]
  {10.1111/j.1365-2966.2010.17190.x}, \href
  {http://adsabs.harvard.edu/abs/2010MNRAS.408..827D} {408, 827}

\bibitem[\protect\citeauthoryear{{Eggenberger}, {Meynet}  \&
  {Maeder}}{{Eggenberger} et~al.}{2002}]{Egg02}
{Eggenberger} P.,  {Meynet} G.,   {Maeder} A.,  2002, \mn@doi [\aap]
  {10.1051/0004-6361:20020262}, \href
  {http://adsabs.harvard.edu/abs/2002A%26A...386..576E} {386, 576}

\bibitem[\protect\citeauthoryear{{Eggleton} \& {Tokovinin}}{{Eggleton} \&
  {Tokovinin}}{2008}]{eggleton2008}
{Eggleton} P.~P.,  {Tokovinin} A.~A.,  2008, \mn@doi [\mnras]
  {10.1111/j.1365-2966.2008.13596.x}, \href
  {http://adsabs.harvard.edu/abs/2008MNRAS.389..869E} {389, 869}

\bibitem[\protect\citeauthoryear{{Folatelli}, {Bersten}, {Kuncarayakti},
  {Benvenuto}, {Maeda}  \& {Nomoto}}{{Folatelli} et~al.}{2015}]{folatelli2015}
{Folatelli} G.,  {Bersten} M.~C.,  {Kuncarayakti} H.,  {Benvenuto} O.~G.,
  {Maeda} K.,   {Nomoto} K.,  2015, preprint, \href
  {http://adsabs.harvard.edu/abs/2015arXiv150901588F} {} (\mn@eprint {arXiv}
  {1509.01588})

\bibitem[\protect\citeauthoryear{{France} et~al.,}{{France}
  et~al.}{2010}]{france2010}
{France} K.,  et~al., 2010, \mn@doi [Science] {10.1126/science.1192134}, \href
  {http://adsabs.harvard.edu/abs/2010Sci...329.1624F} {329, 1624}

\bibitem[\protect\citeauthoryear{{France} et~al.,}{{France}
  et~al.}{2011}]{france2011}
{France} K.,  et~al., 2011, \mn@doi [\apj] {10.1088/0004-637X/743/2/186}, \href
  {http://adsabs.harvard.edu/abs/2011ApJ...743..186F} {743, 186}

\bibitem[\protect\citeauthoryear{{Fransson}, {Cassatella}, {Gilmozzi},
  {Kirshner}, {Panagia}, {Sonneborn}  \& {Wamsteker}}{{Fransson}
  et~al.}{1989}]{fransson1989}
{Fransson} C.,  {Cassatella} A.,  {Gilmozzi} R.,  {Kirshner} R.~P.,  {Panagia}
  N.,  {Sonneborn} G.,   {Wamsteker} W.,  1989, \mn@doi [\apj]
  {10.1086/167022}, \href {http://adsabs.harvard.edu/abs/1989ApJ...336..429F}
  {336, 429}

\bibitem[\protect\citeauthoryear{{Gvaramadze} et~al.,}{{Gvaramadze}
  et~al.}{2015}]{gvaramadze2015}
{Gvaramadze} V.~V.,  et~al., 2015, \mn@doi [\mnras] {10.1093/mnras/stv1995},
  \href {http://adsabs.harvard.edu/abs/2015MNRAS.454..219G} {454, 219}

\bibitem[\protect\citeauthoryear{{Hamuy}, {Suntzeff}, {Gonzalez}  \&
  {Martin}}{{Hamuy} et~al.}{1988}]{hamuy1988}
{Hamuy} M.,  {Suntzeff} N.~B.,  {Gonzalez} R.,   {Martin} G.,  1988, \mn@doi
  [\aj] {10.1086/114613}, \href
  {http://adsabs.harvard.edu/abs/1988AJ.....95...63H} {95, 63}

\bibitem[\protect\citeauthoryear{{Heger} \& {Langer}}{{Heger} \&
  {Langer}}{1998}]{heger1998}
{Heger} A.,  {Langer} N.,  1998, \aap, \href
  {http://adsabs.harvard.edu/abs/1998A\%26A...334..210H} {334, 210}

\bibitem[\protect\citeauthoryear{{Heger} \& {Langer}}{{Heger} \&
  {Langer}}{2000}]{heger2000b}
{Heger} A.,  {Langer} N.,  2000, \mn@doi [\apj] {10.1086/317239}, \href
  {http://adsabs.harvard.edu/abs/2000ApJ...544.1016H} {544, 1016}

\bibitem[\protect\citeauthoryear{{Heger}, {Langer}  \& {Woosley}}{{Heger}
  et~al.}{2000}]{heger2000a}
{Heger} A.,  {Langer} N.,   {Woosley} S.~E.,  2000, \mn@doi [\apj]
  {10.1086/308158}, \href {http://adsabs.harvard.edu/abs/2000ApJ...528..368H}
  {528, 368}

\bibitem[\protect\citeauthoryear{{Heger}, {Woosley}, {Rauscher}, {Hoffman}  \&
  {Boyes}}{{Heger} et~al.}{2002}]{heger2002}
{Heger} A.,  {Woosley} S.~E.,  {Rauscher} T.,  {Hoffman} R.~D.,   {Boyes}
  M.~M.,  2002, \mn@doi [\nar] {10.1016/S1387-6473(02)00184-7}, \href
  {http://adsabs.harvard.edu/abs/2002NewAR..46..463H} {46, 463}

\bibitem[\protect\citeauthoryear{{Heger}, {Woosley}  \& {Spruit}}{{Heger}
  et~al.}{2005}]{heger2005}
{Heger} A.,  {Woosley} S.~E.,   {Spruit} H.~C.,  2005, \mn@doi [\apj]
  {10.1086/429868}, \href {http://adsabs.harvard.edu/abs/2005ApJ...626..350H}
  {626, 350}

\bibitem[\protect\citeauthoryear{{Hendry}, {Smartt}, {Skillman}, {Evans},
  {Trundle}, {Lennon}, {Crowther}  \& {Hunter}}{{Hendry}
  et~al.}{2008}]{hendry2008}
{Hendry} M.~A.,  {Smartt} S.~J.,  {Skillman} E.~D.,  {Evans} C.~J.,  {Trundle}
  C.,  {Lennon} D.~J.,  {Crowther} P.~A.,   {Hunter} I.,  2008, \mn@doi
  [\mnras] {10.1111/j.1365-2966.2008.13347.x}, \href
  {http://adsabs.harvard.edu/abs/2008MNRAS.388.1127H} {388, 1127}

\bibitem[\protect\citeauthoryear{{Hillebrandt} \& {Meyer}}{{Hillebrandt} \&
  {Meyer}}{1989}]{hillebrandt1989}
{Hillebrandt} W.,  {Meyer} F.,  1989, \aap, \href
  {http://adsabs.harvard.edu/abs/1989A%26A...219L...3H} {219, L3}

\bibitem[\protect\citeauthoryear{{Hirschi}, {Meynet}  \& {Maeder}}{{Hirschi}
  et~al.}{2004}]{hirschi2004}
{Hirschi} R.,  {Meynet} G.,   {Maeder} A.,  2004, \mn@doi [\aap]
  {10.1051/0004-6361:20041095}, \href
  {http://adsabs.harvard.edu/abs/2004A\%26A...425..649H} {425, 649}

\bibitem[\protect\citeauthoryear{{Iglesias} \& {Rogers}}{{Iglesias} \&
  {Rogers}}{1996}]{iglesias1996}
{Iglesias} C.~A.,  {Rogers} F.~J.,  1996, \mn@doi [\apj] {10.1086/177381},
  \href {http://adsabs.harvard.edu/abs/1996ApJ...464..943I} {464, 943}

\bibitem[\protect\citeauthoryear{{Ivanova}}{{Ivanova}}{2002}]{ivanovathesis}
{Ivanova} N.,  2002, PhD thesis, University of Oxford

\bibitem[\protect\citeauthoryear{{Ivanova} \& {Podsiadlowski}}{{Ivanova} \&
  {Podsiadlowski}}{2002a}]{ivanova2002a}
{Ivanova} N.,  {Podsiadlowski} P.,  2002a, in {Tout} C.~A.,  {van Hamme} W.,
  eds,  Astronomical Society of the Pacific Conference Series Vol. 279, Exotic
  Stars as Challenges to Evolution. p.~245

\bibitem[\protect\citeauthoryear{{Ivanova} \& {Podsiadlowski}}{{Ivanova} \&
  {Podsiadlowski}}{2002b}]{ivanova2002b}
{Ivanova} N.,  {Podsiadlowski} P.,  2002b, \mn@doi [\apss]
  {10.1023/A:1019553109023}, \href
  {http://adsabs.harvard.edu/abs/2002Ap%26SS.281..191I} {281, 191}

\bibitem[\protect\citeauthoryear{{Ivanova} \& {Podsiadlowski}}{{Ivanova} \&
  {Podsiadlowski}}{2003}]{ivanova2003}
{Ivanova} N.,  {Podsiadlowski} P.,  2003, in {Hillebrandt} W.,  {Leibundgut}
  B.,  eds, From Twilight to Highlight: The Physics of Supernovae. p.~19
  (\mn@eprint {} {astro-ph/0210368}), \mn@doi{10.1007/10828549_3}

\bibitem[\protect\citeauthoryear{{Ivanova}, {Podsiadlowski}  \&
  {Spruit}}{{Ivanova} et~al.}{2002}]{ivanova2002c}
{Ivanova} N.,  {Podsiadlowski} P.,   {Spruit} H.,  2002, \mn@doi [\mnras]
  {10.1046/j.1365-8711.2002.05543.x}, \href
  {http://adsabs.harvard.edu/abs/2002MNRAS.334..819I} {334, 819}

\bibitem[\protect\citeauthoryear{{Justham}, {Podsiadlowski}  \&
  {Vink}}{{Justham} et~al.}{2014}]{justham2014}
{Justham} S.,  {Podsiadlowski} P.,   {Vink} J.~S.,  2014, \mn@doi [\apj]
  {10.1088/0004-637X/796/2/121}, \href
  {http://adsabs.harvard.edu/abs/2014ApJ...796..121J} {796, 121}

\bibitem[\protect\citeauthoryear{{Kleiser} et~al.,}{{Kleiser}
  et~al.}{2011}]{kleiser2011}
{Kleiser} I.~K.~W.,  et~al., 2011, \mn@doi [\mnras]
  {10.1111/j.1365-2966.2011.18708.x}, \href
  {http://adsabs.harvard.edu/abs/2011MNRAS.415..372K} {415, 372}

\bibitem[\protect\citeauthoryear{{Kobulnicky} \& {Fryer}}{{Kobulnicky} \&
  {Fryer}}{2007}]{kobulnicky2007}
{Kobulnicky} H.~A.,  {Fryer} C.~L.,  2007, \mn@doi [\apj] {10.1086/522073},
  \href {http://adsabs.harvard.edu/abs/2007ApJ...670..747K} {670, 747}

\bibitem[\protect\citeauthoryear{{Langer}}{{Langer}}{1991}]{langer1991}
{Langer} N.,  1991, \aap, \href
  {http://adsabs.harvard.edu/abs/1991A%26A...252..669L} {252, 669}

\bibitem[\protect\citeauthoryear{{Lundqvist} \& {Fransson}}{{Lundqvist} \&
  {Fransson}}{1996}]{lundqvist1996}
{Lundqvist} P.,  {Fransson} C.,  1996, \mn@doi [\apj] {10.1086/177380}, \href
  {http://adsabs.harvard.edu/abs/1996ApJ...464..924L} {464, 924}

\bibitem[\protect\citeauthoryear{{Maeder}}{{Maeder}}{1987}]{maeder1987}
{Maeder} A.,  1987, in {Danziger} I.~J.,  ed.,  European Southern Observatory
  Conference and Workshop Proceedings Vol. 26, European Southern Observatory
  Conference and Workshop Proceedings. pp 251--269

\bibitem[\protect\citeauthoryear{{Maran}, {Sonneborn}, {Pun}, {Lundqvist},
  {Iping}  \& {Gull}}{{Maran} et~al.}{2000}]{maran2000}
{Maran} S.~P.,  {Sonneborn} G.,  {Pun} C.~S.~J.,  {Lundqvist} P.,  {Iping}
  R.~C.,   {Gull} T.~R.,  2000, \mn@doi [\apj] {10.1086/317809}, \href
  {http://adsabs.harvard.edu/abs/2000ApJ...545..390M} {545, 390}

\bibitem[\protect\citeauthoryear{{Marigo} \& {Aringer}}{{Marigo} \&
  {Aringer}}{2009}]{marigo2009}
{Marigo} P.,  {Aringer} B.,  2009, \mn@doi [\aap]
  {10.1051/0004-6361/200912598}, \href
  {http://adsabs.harvard.edu/abs/2009A%26A...508.1539M} {508, 1539}

\bibitem[\protect\citeauthoryear{{Mattila}, {Lundqvist}, {Gr{\"o}ningsson},
  {Meikle}, {Stathakis}, {Fransson}  \& {Cannon}}{{Mattila}
  et~al.}{2010}]{mattila2010}
{Mattila} S.,  {Lundqvist} P.,  {Gr{\"o}ningsson} P.,  {Meikle} P.,
  {Stathakis} R.,  {Fransson} C.,   {Cannon} R.,  2010, \mn@doi [\apj]
  {10.1088/0004-637X/717/2/1140}, \href
  {http://adsabs.harvard.edu/abs/2010ApJ...717.1140M} {717, 1140}

\bibitem[\protect\citeauthoryear{{Mazzali} \& {Chugai}}{{Mazzali} \&
  {Chugai}}{1995}]{mazzali1995}
{Mazzali} P.~A.,  {Chugai} N.~N.,  1995, \aap, \href
  {http://adsabs.harvard.edu/abs/1995A\%26A...303..118M} {303, 118}

\bibitem[\protect\citeauthoryear{{Mazzali}, {Lucy}  \& {Butler}}{{Mazzali}
  et~al.}{1992}]{mazzali1992}
{Mazzali} P.~A.,  {Lucy} L.~B.,   {Butler} K.,  1992, \aap, \href
  {http://adsabs.harvard.edu/abs/1992A\%26A...258..399M} {258, 399}

\bibitem[\protect\citeauthoryear{{McCray}}{{McCray}}{1993}]{mccray1993}
{McCray} R.,  1993, \mn@doi [\araa] {10.1146/annurev.aa.31.090193.001135},
  \href {http://adsabs.harvard.edu/abs/1993ARA\%26A..31..175M} {31, 175}

\bibitem[\protect\citeauthoryear{{Melena}, {Massey}, {Morrell}  \&
  {Zangari}}{{Melena} et~al.}{2008}]{melena2008}
{Melena} N.~W.,  {Massey} P.,  {Morrell} N.~I.,   {Zangari} A.~M.,  2008,
  \mn@doi [\aj] {10.1088/0004-6256/135/3/878}, \href
  {http://adsabs.harvard.edu/abs/2008AJ....135..878M} {135, 878}

\bibitem[\protect\citeauthoryear{{Morris} \& {Podsiadlowski}}{{Morris} \&
  {Podsiadlowski}}{2007}]{morris2007}
{Morris} T.,  {Podsiadlowski} P.,  2007, \mn@doi [Science]
  {10.1126/science.1136351}, \href
  {http://adsabs.harvard.edu/abs/2007Sci...315.1103M} {315, 1103}

\bibitem[\protect\citeauthoryear{{Morris} \& {Podsiadlowski}}{{Morris} \&
  {Podsiadlowski}}{2009}]{morris2009}
{Morris} T.,  {Podsiadlowski} P.,  2009, \mn@doi [\mnras]
  {10.1111/j.1365-2966.2009.15114.x}, \href
  {http://adsabs.harvard.edu/abs/2009MNRAS.399..515M} {399, 515}

\bibitem[\protect\citeauthoryear{{Nieuwenhuijzen} \& {de
  Jager}}{{Nieuwenhuijzen} \& {de Jager}}{1990}]{nieuwenhuijzen1990}
{Nieuwenhuijzen} H.,  {de Jager} C.,  1990, \aap, \href
  {http://adsabs.harvard.edu/abs/1990A\%26A...231..134N} {231, 134}

\bibitem[\protect\citeauthoryear{{Nomoto}, {Shigeyama}, {Kumaga}  \&
  {Hashimoto}}{{Nomoto} et~al.}{1988}]{nomoto1988}
{Nomoto} K.,  {Shigeyama} T.,  {Kumaga} S.,   {Hashimoto} M.-A.,  1988,
  Proceedings of the Astronomical Society of Australia, \href
  {http://adsabs.harvard.edu/abs/1988PASAu...7..490N} {7, 490}

\bibitem[\protect\citeauthoryear{{Panagia}, {Scuderi}, {Gilmozzi}, {Challis},
  {Garnavich}  \& {Kirshner}}{{Panagia} et~al.}{1996}]{panagia1996}
{Panagia} N.,  {Scuderi} S.,  {Gilmozzi} R.,  {Challis} P.~M.,  {Garnavich}
  P.~M.,   {Kirshner} R.~P.,  1996, \mn@doi [\apjl] {10.1086/309930}, \href
  {http://adsabs.harvard.edu/abs/1996ApJ...459L..17P} {459, L17}

\bibitem[\protect\citeauthoryear{{Pastorello} et~al.,}{{Pastorello}
  et~al.}{2012}]{pastorello2012}
{Pastorello} A.,  et~al., 2012, \mn@doi [\aap] {10.1051/0004-6361/201118112},
  \href {http://adsabs.harvard.edu/abs/2012A%26A...537A.141P} {537, A141}

\bibitem[\protect\citeauthoryear{{Petermann}, {Langer}, {Castro}  \&
  {Fossati}}{{Petermann} et~al.}{2015}]{petermann2015}
{Petermann} I.,  {Langer} N.,  {Castro} N.,   {Fossati} L.,  2015, \mn@doi
  [\aap] {10.1051/0004-6361/201526302}, \href
  {http://adsabs.harvard.edu/abs/2015A%26A...584A..54P} {584, A54}

\bibitem[\protect\citeauthoryear{{Podsiadlowski}}{{Podsiadlowski}}{1992}]{pods1992b}
{Podsiadlowski} P.,  1992, \mn@doi [\pasp] {10.1086/133043}, \href
  {http://adsabs.harvard.edu/abs/1992PASP..104..717P} {104, 717}

\bibitem[\protect\citeauthoryear{{Podsiadlowski} \& {Joss}}{{Podsiadlowski} \&
  {Joss}}{1989}]{pods1989}
{Podsiadlowski} P.,  {Joss} P.~C.,  1989, \mn@doi [\nat] {10.1038/338401a0},
  \href {http://adsabs.harvard.edu/abs/1989Natur.338..401P} {338, 401}

\bibitem[\protect\citeauthoryear{{Podsiadlowski}, {Joss}  \&
  {Rappaport}}{{Podsiadlowski} et~al.}{1990}]{pods1990}
{Podsiadlowski} P.,  {Joss} P.~C.,   {Rappaport} S.,  1990, \aap, \href
  {http://adsabs.harvard.edu/abs/1990A%26A...227L...9P} {227, L9}

\bibitem[\protect\citeauthoryear{{Podsiadlowski}, {Joss}  \&
  {Hsu}}{{Podsiadlowski} et~al.}{1992}]{pods1992a}
{Podsiadlowski} P.,  {Joss} P.~C.,   {Hsu} J.~J.~L.,  1992, \mn@doi [\apj]
  {10.1086/171341}, \href {http://adsabs.harvard.edu/abs/1992ApJ...391..246P}
  {391, 246}

\bibitem[\protect\citeauthoryear{{Podsiadlowski}, {Morris}  \&
  {Ivanova}}{{Podsiadlowski} et~al.}{2006}]{pods2006}
{Podsiadlowski} P.,  {Morris} T.~S.,   {Ivanova} N.,  2006, in {Kraus} M.,
  {Miroshnichenko} A.~S.,  eds,  Astronomical Society of the Pacific Conference
  Series Vol. 355, Stars with the B[e] Phenomenon. p.~259

\bibitem[\protect\citeauthoryear{{Podsiadlowski}, {Morris}  \&
  {Ivanova}}{{Podsiadlowski} et~al.}{2007}]{pods2007}
{Podsiadlowski} P.,  {Morris} T.~S.,   {Ivanova} N.,  2007, in {Immler} S.,
  {Weiler} K.,   {McCray} R.,  eds,  American Institute of Physics Conference
  Series Vol. 937, Supernova 1987A: 20 Years After: Supernovae and Gamma-Ray
  Bursters. pp 125--133, \mn@doi{10.1063/1.3682893}

\bibitem[\protect\citeauthoryear{{Popova}, {Tutukov}  \& {Yungelson}}{{Popova}
  et~al.}{1982}]{popova1982}
{Popova} E.~I.,  {Tutukov} A.~V.,   {Yungelson} L.~R.,  1982, \mn@doi [\apss]
  {10.1007/BF00648989}, \href
  {http://adsabs.harvard.edu/abs/1982Ap%26SS..88...55P} {88, 55}

\bibitem[\protect\citeauthoryear{{Potekhin}, {Chabrier}, {Lai}, {Ho}  \& {van
  Adelsberg}}{{Potekhin} et~al.}{2006}]{potekhin2006}
{Potekhin} A.~Y.,  {Chabrier} G.,  {Lai} D.,  {Ho} W.~C.~G.,   {van Adelsberg}
  M.,  2006, \mn@doi [Journal of Physics A Mathematical General]
  {10.1088/0305-4470/39/17/S21}, \href
  {http://adsabs.harvard.edu/abs/2006JPhA...39.4453P} {39, 4453}

\bibitem[\protect\citeauthoryear{{Rauscher}, {Heger}, {Hoffman}  \&
  {Woosley}}{{Rauscher} et~al.}{2002}]{rauscher2002}
{Rauscher} T.,  {Heger} A.,  {Hoffman} R.~D.,   {Woosley} S.~E.,  2002, \mn@doi
  [\apj] {10.1086/341728}, \href
  {http://adsabs.harvard.edu/abs/2002ApJ...576..323R} {576, 323}

\bibitem[\protect\citeauthoryear{{Saio}, {Nomoto}  \& {Kato}}{{Saio}
  et~al.}{1988}]{saio1988}
{Saio} H.,  {Nomoto} K.,   {Kato} M.,  1988, \mn@doi [\apj] {10.1086/166565},
  \href {http://adsabs.harvard.edu/abs/1988ApJ...331..388S} {331, 388}

\bibitem[\protect\citeauthoryear{{Sana} et~al.,}{{Sana}
  et~al.}{2012}]{sana2012}
{Sana} H.,  et~al., 2012, \mn@doi [Science] {10.1126/science.1223344}, \href
  {http://adsabs.harvard.edu/abs/2012Sci...337..444S} {337, 444}

\bibitem[\protect\citeauthoryear{{Sana} et~al.,}{{Sana}
  et~al.}{2013}]{sana2013}
{Sana} H.,  et~al., 2013, \mn@doi [\aap] {10.1051/0004-6361/201219621}, \href
  {http://adsabs.harvard.edu/abs/2013A%26A...550A.107S} {550, A107}

\bibitem[\protect\citeauthoryear{{Smartt}}{{Smartt}}{2009}]{smartt2009a}
{Smartt} S.~J.,  2009, \mn@doi [\araa] {10.1146/annurev-astro-082708-101737},
  \href {http://adsabs.harvard.edu/abs/2009ARA%26A..47...63S} {47, 63}

\bibitem[\protect\citeauthoryear{{Smartt}, {Lennon}, {Kudritzki}, {Rosales},
  {Ryans}  \& {Wright}}{{Smartt} et~al.}{2002}]{smartt2002}
{Smartt} S.~J.,  {Lennon} D.~J.,  {Kudritzki} R.~P.,  {Rosales} F.,  {Ryans}
  R.~S.~I.,   {Wright} N.,  2002, \mn@doi [\aap] {10.1051/0004-6361:20020829},
  \href {http://adsabs.harvard.edu/abs/2002A%26A...391..979S} {391, 979}

\bibitem[\protect\citeauthoryear{{Smartt}, {Eldridge}, {Crockett}  \&
  {Maund}}{{Smartt} et~al.}{2009}]{smartt2009b}
{Smartt} S.~J.,  {Eldridge} J.~J.,  {Crockett} R.~M.,   {Maund} J.~R.,  2009,
  \mn@doi [\mnras] {10.1111/j.1365-2966.2009.14506.x}, \href
  {http://adsabs.harvard.edu/abs/2009MNRAS.395.1409S} {395, 1409}

\bibitem[\protect\citeauthoryear{{Smith}, {Nota}, {Pasquali}, {Leitherer},
  {Clampin}  \& {Crowther}}{{Smith} et~al.}{1998}]{smith1998}
{Smith} L.~J.,  {Nota} A.,  {Pasquali} A.,  {Leitherer} C.,  {Clampin} M.,
  {Crowther} P.~A.,  1998, \mn@doi [\apj] {10.1086/305980}, \href
  {http://adsabs.harvard.edu/abs/1998ApJ...503..278S} {503, 278}

\bibitem[\protect\citeauthoryear{{Smith}, {Arnett}, {Bally}, {Ginsburg}  \&
  {Filippenko}}{{Smith} et~al.}{2013}]{smith2013}
{Smith} N.,  {Arnett} W.~D.,  {Bally} J.,  {Ginsburg} A.,   {Filippenko} A.~V.,
   2013, \mn@doi [\mnras] {10.1093/mnras/sts418}, \href
  {http://adsabs.harvard.edu/abs/2013MNRAS.429.1324S} {429, 1324}

\bibitem[\protect\citeauthoryear{{Sonneborn}, {Fransson}, {Lundqvist},
  {Cassatella}, {Gilmozzi}, {Kirshner}, {Panagia}  \& {Wamsteker}}{{Sonneborn}
  et~al.}{1997}]{sonneborn1997}
{Sonneborn} G.,  {Fransson} C.,  {Lundqvist} P.~A.,  {Cassatella} A.,
  {Gilmozzi} R.,  {Kirshner} R.~P.,  {Panagia} N.,   {Wamsteker} W.,  1997,
  \apj, \href {http://adsabs.harvard.edu/abs/1997ApJ...477..848S} {477, 848}

\bibitem[\protect\citeauthoryear{{Sugerman}, {Crotts}, {Kunkel}, {Heathcote}
  \& {Lawrence}}{{Sugerman} et~al.}{2005a}]{sugerman2005a}
{Sugerman} B.~E.~K.,  {Crotts} A.~P.~S.,  {Kunkel} W.~E.,  {Heathcote} S.~R.,
  {Lawrence} S.~S.,  2005a, \mn@doi [\apjs] {10.1086/430408}, \href
  {http://adsabs.harvard.edu/abs/2005ApJS..159...60S} {159, 60}

\bibitem[\protect\citeauthoryear{{Sugerman}, {Crotts}, {Kunkel}, {Heathcote}
  \& {Lawrence}}{{Sugerman} et~al.}{2005b}]{sugerman2005b}
{Sugerman} B.~E.~K.,  {Crotts} A.~P.~S.,  {Kunkel} W.~E.,  {Heathcote} S.~R.,
  {Lawrence} S.~S.,  2005b, \mn@doi [\apj] {10.1086/430396}, \href
  {http://adsabs.harvard.edu/abs/2005ApJ...627..888S} {627, 888}

\bibitem[\protect\citeauthoryear{{Taddia} et~al.,}{{Taddia}
  et~al.}{2013}]{taddia2013}
{Taddia} F.,  et~al., 2013, \mn@doi [\aap] {10.1051/0004-6361/201322276}, \href
  {http://adsabs.harvard.edu/abs/2013A%26A...558A.143T} {558, A143}

\bibitem[\protect\citeauthoryear{{Tutukov}, {Yungelson}  \& {Iben}}{{Tutukov}
  et~al.}{1992}]{tutukov1992}
{Tutukov} A.~V.,  {Yungelson} L.~R.,   {Iben} Jr. I.,  1992, \mn@doi [\apj]
  {10.1086/171005}, \href {http://adsabs.harvard.edu/abs/1992ApJ...386..197T}
  {386, 197}

\bibitem[\protect\citeauthoryear{{Ugliano}, {Janka}, {Marek}  \&
  {Arcones}}{{Ugliano} et~al.}{2012}]{ugliano2012}
{Ugliano} M.,  {Janka} H.-T.,  {Marek} A.,   {Arcones} A.,  2012, \mn@doi
  [\apj] {10.1088/0004-637X/757/1/69}, \href
  {http://adsabs.harvard.edu/abs/2012ApJ...757...69U} {757, 69}

\bibitem[\protect\citeauthoryear{{Utrobin} \& {Chugai}}{{Utrobin} \&
  {Chugai}}{2005}]{utrobin2005}
{Utrobin} V.~P.,  {Chugai} N.~N.,  2005, \mn@doi [\aap]
  {10.1051/0004-6361:20042599}, \href
  {http://adsabs.harvard.edu/abs/2005A%26A...441..271U} {441, 271}

\bibitem[\protect\citeauthoryear{{Utrobin}, {Wongwathanarat}, {Janka}  \&
  {M{\"u}ller}}{{Utrobin} et~al.}{2015}]{utrobin2015}
{Utrobin} V.~P.,  {Wongwathanarat} A.,  {Janka} H.-T.,   {M{\"u}ller} E.,
  2015, \mn@doi [\aap] {10.1051/0004-6361/201425513}, \href
  {http://adsabs.harvard.edu/abs/2015A%26A...581A..40U} {581, A40}

\bibitem[\protect\citeauthoryear{{Vanbeveren}, {Mennekens}, {Van Rensbergen}
  \& {De Loore}}{{Vanbeveren} et~al.}{2013}]{vanbeveren2013}
{Vanbeveren} D.,  {Mennekens} N.,  {Van Rensbergen} W.,   {De Loore} C.,  2013,
  \mn@doi [\aap] {10.1051/0004-6361/201321072}, \href
  {http://adsabs.harvard.edu/abs/2013A%26A...552A.105V} {552, A105}

\bibitem[\protect\citeauthoryear{{Walborn}, {Lasker}, {Laidler}  \&
  {Chu}}{{Walborn} et~al.}{1987}]{walborn1987}
{Walborn} N.~R.,  {Lasker} B.~M.,  {Laidler} V.~G.,   {Chu} Y.-H.,  1987,
  \mn@doi [\apjl] {10.1086/185002}, \href
  {http://adsabs.harvard.edu/abs/1987ApJ...321L..41W} {321, L41}

\bibitem[\protect\citeauthoryear{{Walborn}, {Prevot}, {Prevot}, {Wamsteker},
  {Gonzalez}, {Gilmozzi}  \& {Fitzpatrick}}{{Walborn}
  et~al.}{1989}]{walborn1989}
{Walborn} N.~R.,  {Prevot} M.~L.,  {Prevot} L.,  {Wamsteker} W.,  {Gonzalez}
  R.,  {Gilmozzi} R.,   {Fitzpatrick} E.~L.,  1989, \aap, \href
  {http://adsabs.harvard.edu/abs/1989A%26A...219..229W} {219, 229}

\bibitem[\protect\citeauthoryear{{Wampler}, {Wang}, {Baade}, {Banse},
  {D'Odorico}, {Gouiffes}  \& {Tarenghi}}{{Wampler} et~al.}{1990}]{wampler1990}
{Wampler} E.~J.,  {Wang} L.,  {Baade} D.,  {Banse} K.,  {D'Odorico} S.,
  {Gouiffes} C.,   {Tarenghi} M.,  1990, \mn@doi [\apjl] {10.1086/185836},
  \href {http://adsabs.harvard.edu/abs/1990ApJ...362L..13W} {362, L13}

\bibitem[\protect\citeauthoryear{{Weiss}, {Hillebrandt}  \& {Truran}}{{Weiss}
  et~al.}{1988}]{weiss1988}
{Weiss} A.,  {Hillebrandt} W.,   {Truran} J.~W.,  1988, \aap, \href
  {http://adsabs.harvard.edu/abs/1988A%26A...197L..11W} {197, L11}

\bibitem[\protect\citeauthoryear{{West}, {Lauberts}, {Schuster}  \&
  {Jorgensen}}{{West} et~al.}{1987}]{west1987}
{West} R.~M.,  {Lauberts} A.,  {Schuster} H.-E.,   {Jorgensen} H.~E.,  1987,
  \aap, \href {http://adsabs.harvard.edu/abs/1987A%26A...177L...1W} {177, L1}

\bibitem[\protect\citeauthoryear{{Wood}}{{Wood}}{1988}]{wood1988}
{Wood} P.~R.,  1988, Proceedings of the Astronomical Society of Australia,
  \href {http://adsabs.harvard.edu/abs/1988PASAu...7..386W} {7, 386}

\bibitem[\protect\citeauthoryear{{Woosley}}{{Woosley}}{1988}]{woosley1988a}
{Woosley} S.~E.,  1988, \mn@doi [\apj] {10.1086/166468}, \href
  {http://adsabs.harvard.edu/abs/1988ApJ...330..218W} {330, 218}

\bibitem[\protect\citeauthoryear{{Woosley} \& {Heger}}{{Woosley} \&
  {Heger}}{2007}]{woosley2007a}
{Woosley} S.~E.,  {Heger} A.,  2007, \mn@doi [\physrep]
  {10.1016/j.physrep.2007.02.009}, \href
  {http://adsabs.harvard.edu/abs/2007PhR...442..269W} {442, 269}

\bibitem[\protect\citeauthoryear{{Woosley}, {Pinto}  \& {Weaver}}{{Woosley}
  et~al.}{1988}]{woosley1988b}
{Woosley} S.~E.,  {Pinto} P.~A.,   {Weaver} T.~A.,  1988, Proceedings of the
  Astronomical Society of Australia, \href
  {http://adsabs.harvard.edu/abs/1988PASAu...7..355W} {7, 355}

\bibitem[\protect\citeauthoryear{{Woosley}, {Heger}, {Weaver}  \&
  {Langer}}{{Woosley} et~al.}{1997}]{woosley1997}
{Woosley} S.~E.,  {Heger} A.,  {Weaver} T.~A.,   {Langer} N.,  1997, ArXiv
  Astrophysics e-prints, \href
  {http://adsabs.harvard.edu/abs/1997astro.ph..5146W} {}

\bibitem[\protect\citeauthoryear{{Woosley}, {Heger}  \& {Weaver}}{{Woosley}
  et~al.}{2002}]{woosley2002}
{Woosley} S.~E.,  {Heger} A.,   {Weaver} T.~A.,  2002, \mn@doi [Reviews of
  Modern Physics] {10.1103/RevModPhys.74.1015}, \href
  {http://adsabs.harvard.edu/abs/2002RvMP...74.1015W} {74, 1015}

\bibitem[\protect\citeauthoryear{{de Mink}, {Pols}  \& {Yoon}}{{de Mink}
  et~al.}{2008}]{demink2008}
{de Mink} S.~E.,  {Pols} O.~R.,   {Yoon} S.-C.,  2008, in {O'Shea} B.~W.,
  {Heger} A.,  eds,  American Institute of Physics Conference Series Vol. 990,
  First Stars III. pp 230--232 (\mn@eprint {arXiv} {0710.1010}),
  \mn@doi{10.1063/1.2905549}

\makeatother
\end{thebibliography}
\bsp	
\label{lastpage}
\end{document}